\documentclass[preprint,twocolumn]{aastex631}
\usepackage{subfigure}
\usepackage{color}
\usepackage{xcolor}
\usepackage{amsmath}
\usepackage{mathrsfs}
\usepackage{multirow}
\usepackage{soul}
\usepackage{amssymb}
\usepackage{hyperref}
\usepackage{threeparttable}
\everymath{\displaystyle} 

\definecolor{ultramarine}{rgb}{0.8, 0.1, 0.4}
\defcitealias{2025ApJ...980..210Z}{Paper II}
\defcitealias{ZA24a}{Paper I}

\def\be{\begin{equation}}
\def\ee{\end{equation}}
\def\bd{\begin{displaymath}}
\def\ed{\end{displaymath}}
\def\ba{\begin{aligned}}
\def\ea{\end{aligned}}

\def\nms{\mathsurround=0pt}

\def\oversim#1#2{\lower 4pt\vbox{\baselineskip 0pt \lineskip 1pt
    \ialign{$\nms#1\hfil##\hfil$\crcr#2\crcr\sim\crcr}}}

\def\bh{M_{\bullet}}

\def\msun{M_{\odot}}

\def\kms{{\rm \,km\,s^{-1}}}

\def\pyr{{\rm yr}^{-1}}

\def\GNC{\texttt{GNC}}

\begin{document}

\title{Stellar Collisions from Self-consistent Stellar Dynamics Around Growing Supermassive black Holes}
\author{Fupeng Zhang}
\correspondingauthor{Fupeng Zhang}
\email{zhangfupeng@gzhu.edu.cn}
\affiliation{School of Physics and Materials Science, Guangzhou
University, Guangzhou 510006, People's Republic of China}
\affiliation{Key Laboratory for Astronomical Observation and Technology of Guangzhou, 510006 Guangzhou, People's Republic of China}
\affiliation{Astronomy Science and Technology Research Laboratory of Department of Education of Guangdong Province, Guangzhou 510006, People's Republic of China}
\author{Pau Amaro Seoane}
\affiliation{Universitat Politècnica de València, Spain}
\affiliation{
Max-Planck-Institute for Extraterrestrial Physics, Garching, Germany}
\affiliation{
Kavli Institute for Astronomy and Astrophysics, Beijing, China}
\begin{abstract}
The centers of galaxies harbor the densest stellar environments, where a massive black hole (MBH) accelerates stars to such high velocities that direct collisions can result in high-energetic phenomena, such as gravitational wave sources, kilonovae, and supernova-like transients. These collisions can reshape the cluster's density profile and release gas that can be subsequently accreted by the MBH. However, the evolving rates of such phenomena from self-consistent dynamics around mass-growing MBHs remain largely unexplored. In this work, we simulate nuclear star clusters (NSCs) across a range of masses and density profiles by employing the \GNC~Monte Carlo code, that self-consistently models stellar dynamics and the subsequent accretion of released gas. We find that stellar collisions flatten the density cusp in the innermost regions ($r \lesssim 10^{-3}-10^{-2}$ pc) within $\sim 0.1-1$ Gyr. While high initial collision rates in steep cusps quickly decline due to stellar depletion, the interplay between collisions and MBH growth is important only in massive NSCs ($M_\star \sim 10^9\msun$). As MBH grows, increased stellar velocities shift the balance toward destructive collisions of which relative velocities can be $\gtrsim 2500\kms$. Consequently, present-day destructive collision rates in massive clusters remain high ($10^{-4}\sim 10^{-3}\pyr$), whereas they are smaller in Milky Way-like NSCs or negligible in smaller NSCs. Our results highlight a crucial synergy between stellar dynamics and MBH growth, identifying massive galaxies as prime targets for observing transients from destructive stellar collisions.
\end{abstract}


\section{Introduction}

Nuclear star clusters (NSCs) represent the densest stellar systems observed in the Universe, commonly residing at the dynamical centers of galaxies across the Hubble sequence~\citep[for a review, see][]{2020A&ARv..28....4N}. Many, if not all, of these NSCs host a massive black hole (MBH). The extreme densities and the deep potential well dominated by the MBH create a unique astrophysical laboratory where a rich variety of stellar dynamical processes occur at accelerated rates.

The high stellar number densities, coupled with the orbital velocities induced by the MBH, facilitate frequent gravitational interactions. Stars on low angular momentum orbits can be scattered into the loss cone, resulting in their tidal disruption~\citep{1988Natur.333..523R} or direct swallowing by the MBH. Furthermore, the inspiral of compact objects into the MBH generates gravitational waves (GWs), known as Extreme Mass Ratio Inspirals (EMRIs), which are key targets for future space-borne GW detectors~\citep{2005ApJ...629..362H,2006ApJ...645L.133H,2006ApJ...645.1152H, AmaroSeoane17, 2018LRR....21....4A, 2020GReGr..52...81B, 2023LRR....26....2A}.

Among the most dramatic outcomes of this dense environment are direct physical collisions between stars. The possibility that stellar collisions could significantly influence the evolution of galactic nuclei and fuel the central MBH has long been recognized~\citep{1966ApJ...143..400S, 1970ApJ...162..791S,1991ApJ...370...60M,2002A&A...394..345F}. Close to the MBH, the Keplerian velocities can vastly exceed the escape velocities of the stars themselves. Consequently, collisions are often hypervelocity events, releasing tremendous amounts of energy and potentially disrupting the stars involved~\citep[e.g.,][]{2022ApJS..260....2R,2026ApJ..1000..162R}.

Stellar collisions play a multifaceted role in the evolution of NSCs. They can alter the stellar mass function, potentially leading to the formation of massive stars through repeated mergers~\citep{2002ApJ...576..899P, 2023MNRAS.523.5439R} or even the runaway growth of an intermediate-mass black hole (IMBH) seed~\citep{1990ApJ...356..483Q,2017MNRAS.467.4180S}. Collisions can also modify the spatial distribution of stars~\citep{2025A&A...695A..98A, 2023ApJ...955...30R,2024ApJ...963L..17R}. For instance, the observed depletion of red giants (RGs) in the Galactic Center has been partly attributed to collisions or tidal stripping~\citep{2023MNRAS.526.3688B}.

Recently, \citet{2023ApJ...947....8A} emphasized the significance of destructive stellar collisions as high-energy astrophysical phenomena. They derived analytical estimates for the rates of these events, suggesting they could mimic supernovae or even TDEs. Moreover, they highlighted a particularly interesting channel: collisions between two RGs. If the collision is sufficiently close, the degenerate cores of the RGs might survive and form a tight binary within the common envelope of the merged stellar debris. The subsequent evolution of this binary, driven by GW emission, could lead to a merger observable by GW detectors, potentially accompanied by a kilonova-like electromagnetic transient. The analytical work suggested that the collision rates are high enough to rapidly deplete the stellar population in the innermost regions, thereby regulating the collision rate itself.

Despite significant progress in modeling the dynamics of NSCs and stellar collisions~\citep{1971ApJ...164..399S,SM78,2023ApJ...955...30R,2024ApJ...963L..17R,1980ApJ...239..685M,1983ApJ...268..565D,1987ApJ...321..199Q,1991ApJ...370...60M, 2023ApJ...947....8A,2023ApJ...955...30R,2024ApJ...963L..17R}, most studies have relied on analytical estimations or numerical simulations that often assume a static stellar potential and a fixed mass of MBH. However, the very processes driving the dynamics—TDEs, direct swallowing of stars, and the accretion of gas released during stellar collisions—inevitably lead to both the evolution of the density profile of the cluster and the growth of the MBH. The evolution of both the cluster and the MBH, in turn, alters the gravitational potential and the dynamics of the surrounding stars, creating a complex feedback loop. 

Incorporating all of the above complexities is numerically challenging. Currently, only a few studies attempt to address collision processes including these complexities. Early work by \citet{1983ApJ...268..565D} and ~\citet{1991ApJ...370...60M} adopted one-dimensional Fokker-Plank method to study the evolution of the galactic nuclei incorporate tidal disruptions, stellar collisions and accretion of MBHs. However, their treatments of collision terms relied on complex analytical arguments, making it difficult to extend to include further layers of complexity. 
We notice that \citet{2002A&A...394..345F} developed a sophisticated Monte-Carlo method based on H{\'e}non's scheme~\citep{1961AnAp...24..369H}, ME(SSY), which can account most of the above processes. Nevertheless, they focused mainly on the accretion process of MBH and do not specifically address the rates and properties of destructive collisions evolve together with the growing MBH and the changing stellar cusp. Thus, the long-term, self-consistent co-evolution of NSCs and their central MBHs, incorporating the detailed interplay of stellar collisions and accretion, remains a largely uncharted territory.

In this study, we explore this crucial interaction by investigating the dynamics of stellar collisions around self-consistent stellar dynamics and growing MBHs. We utilize the Gravitational N-body Code (\GNC)~\citep{ZA24a, 2025ApJ...980..210Z,2026ApJ...999..224Z}, a Monte Carlo method that is capable of simulating the two-dimensional (energy and angular momentum) self-consistent evolution of stellar relaxation in multi-component NSCs. We have updated \GNC~to include a detailed treatment of stellar collisions, distinguishing between destructive and non-destructive events, and tracking the release of gaseous debris. Crucially, we allow the MBH to grow by accreting this debris, as well as through TDEs and the direct swallowing of stars. This allows us to study the evolving rates and properties of high-energy collisional phenomena in a self-consistent framework where the MBH mass evolves dynamically.

The paper is organized as follows. In Section 2, we describe the updates to \GNC~to include stellar collisions and the verification tests performed. In Section 3, we present the results of our simulations, analyzing the dynamics of stellar collisions first with a fixed MBH mass and then with a growing MBH, comparing the outcomes across different NSC models. Finally, in Section 4, we discuss the implications of our findings and present our conclusions.

\section{The method}
\label{sec:method}

\begin{figure}
    \center
    \includegraphics[scale=0.65]{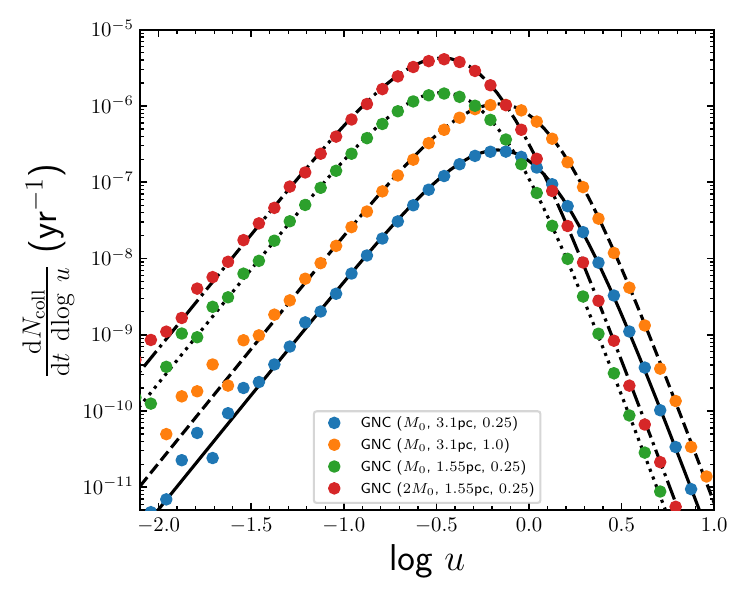}
    \caption{
Rate of collisions per unit logarithmic separation of radius for a Plummer cluster. {Numbers in parentheses of the legend are $M$,  $r_a$ and $\zeta$, respectively, where $M$ is the cluster mass in unit of $M_0=4\times10^6\msun$, $r_a$ is the Plummer scale radius, and $\zeta$ is the parameter controlling the closest-approach distance of the collision pair, as defined in Equation~\ref{eq:gammaij}. The horizontal axis $u = r / r_a$ is the dimensionless radius. The cluster consists of $1\,\msun$ stars.} The black lines represents the theoretical expectation from Equation~\ref{eq:collision_rate_plummer}. The filled colored circles are the instantaneous rate obtained by \GNC. The agreement between the numerical results and the theoretical expectation validates our implementation of stellar collisions in \GNC.
    }
\label{fig:plummer_test}
\end{figure}

\begin{figure}
    \center
    \includegraphics[scale=0.65]{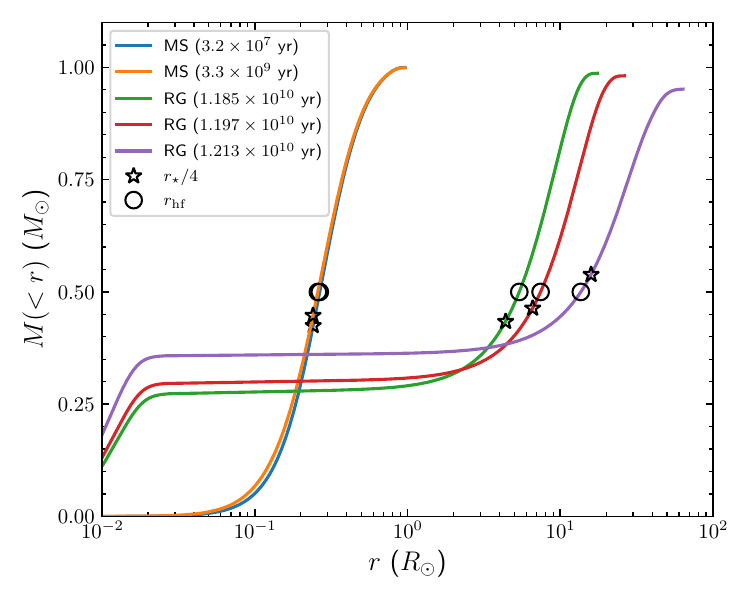}
    \caption{
{Evolution of mass-radius relation for a $1\msun$ star from MESA. $M(<r)$ is the enclosed mass within a distance $r$ from the center of star. ``MS'' and ``RG'' denote the main-sequence and red-giant phases, respectively. The numbers in brackets of the legend are the ages of the star. The open circle and star symbols are the half-mass radius and $r_\star/4$, respectively, where $r_\star$ is the stellar radius.}
    }
\label{fig:mr}
\end{figure}

Our method is based on \GNC~\citep{ZA24a,2025ApJ...980..210Z, 2026ApJ...999..224Z}, a Monte Carlo approach capable of considering the dynamical evolution of a NSC composed of multiple mass components and different stellar species. Our method accounts for two-body relaxation by solving the two-dimensional Fokker-Planck equation in the space of energy and angular momentum~\citep[][Here after Paper I]{ZA24a}. Additionally, \GNC~achieves self-consistent solutions for the evolution of the NSC by solving the Poisson equation and employing adiabatic invariant theory~\citep[][Here after Paper II]{2025ApJ...980..210Z}. \GNC~incorporates loss cone effects, which lead to the tidal disruption of stars or direct swallowing of objects by the massive black hole. In this study, we extend \GNC~in order to address stellar collisions. The details and the tests of the new added numerical scheme are shown in the following sections.

\subsection{Integrating the numerical simulation of collision events}
The numerical scheme to do so follows a similar approach to that of~\citet{2002A&A...394..345F}. However, since the Monte Carlo method in \GNC~is fundamentally distinct from that employed by~\citet{2002A&A...394..345F}, certain modifications are necessary to obtain accurate statistics for collision events. For instance, the true number of stellar objects represented by each particle in our simulation differs, as the weighting of each particle is not uniform, unlike the approach adopted in their method. We elaborate now on the particular aspects of our algorithm.

During each iteration of \GNC~(spanning a time interval $\Delta t$), particles evolve according to the stellar dynamics of individual stars within the cluster, as outlined in Paper II. Subsequently, for individual stars, we assign a random position $r$ and velocity $v$ based on their current orbital energy $E$ and angular momentum $J$ by using a method similar to that described by~\citet{1971Ap&SS..14..151H}. Then, the total velocity of the particle is given by $v=\sqrt{2(\phi(r)-E)}$, where $\phi=\phi_\star(r)+G\bh/r$ is the gravitational potential, and the transverse velocity by $v_t=J/r$ for a given position $r$. {Note that we use the convention of potential and energy being positive for bound orbits (See more details in Section 2.2 of~\citetalias{2025ApJ...980..210Z})}.

Next, we rank all samples by their values of $r$ in ascending order. More specifically, using $i$ and $j$ to indicate the indices of the {pair of samples}, $w_i$ ($w_j$) represents the copy of the weights $\mathcal{W}_i$ ($\mathcal{W}_j$) of particle $i$ ($j$)\footnote{Note that the weight of a particle corresponds to the true number of stellar objects represented by the particle.}. Hence, the calculation of the collision process is performed in the following fashion (starting with $i=1$, $j=2$): We calculate the rate of collision between the two particles,

\be
R_{ij}=\pi b^2_{\rm max} n(r)v_{\rm rel}= 
\pi d^2_{\rm max}n(r)v_{\rm rel}\left(1+\frac{2m_t G}{d_{\rm max}v^2_{\rm rel}}\right),
\label{eq:gammaij}
\ee
\noindent
where $m_t=m_i+m_j$, {$b_{\rm max}$ is the maximum impact parameter, $d_{\rm max}=\zeta(r_{\star,i}+r_{\star,j})$, $v_{\rm rel}=|\vec v_i-\vec v_j|$, and $m_i$ ($m_j$), $r_{\star,i}$ ($r_{\star,j}$), and $\vec v_i$ ($\vec v_j$) are the mass, radius, and velocity of particle $i$ ($j$), respectively.} Here, $n(r)$ is the density of all stellar objects at position $r$, {$d_{\rm max}$ is the maximum closest-approach distance, and $\zeta$ is a parameter controlling the value of $d_{\rm max}$. The position $r=(r_i+r_j)/2$ is the mean distance of two particles from the MBH}. We assume that the orientations of the velocities of the two particles are both randomly distributed. We denote $v_{i,t}$ ($v_{j,t}$) and $v_{i,r}$ ($v_{j,r}$) as the transverse and radial velocities of particle $i$ ($j$), respectively. Then, we set $\vec v_i=(v_{i,t},0,v_{i,r})$ in Cartesian coordinates $(x,y,z)$ and $\vec v_j=(v_{j,t}\cos\xi,v_{j,t}\sin\xi,\pm v_{j,r})$, where $\xi$ is a uniform variable in $(0,\pi)$. {The impact parameter for each collision is selected according to 
the probability density function $p(b)\propto b$ for $b\in(0,b_{\rm max})$~\citep[See  Equation 20
of][]{2002A&A...394..345F}. Then the corresponding closest-approach distance is given by
$d=\sqrt{b^2+F^2}-F$, where $F=G m_t/v_{\rm rel}^2$.
}

Given $\Delta t$, the duration of the simulation in the current iteration, the probability of collision between the pair is given by $R_{ij}\Delta t$. We ensure that the weights of the two particles are equal for this pair of collisions by splitting the particle with the larger weight into two parts. For instance, if $w_i>w_j$, we split the $i$-th particle into two parts, with weights, $w_j$ and $w_i-w_j$ correspondingly. The first part is used to evaluate the collision. Thus, the expected number of collisions for this pair of samples is given by $N_{\rm col}(\Delta t)={\rm min}(w_i,w_j)R_{ij}\Delta t$. 

We then update the weights as follows: if $w_i\geq w_j$, set $w_i\rightarrow w_i-w_j$ and $w_j\rightarrow 0$; if $w_i<w_j$, set $w_i\rightarrow 0$ and $w_j\rightarrow w_j-w_i$. After that, if $w_i>0$, $i$ remains unchanged; otherwise, $i$ and $w_i$ are set to the index and a copy of the weight of the next particle with a non-zero weight, respectively. Similarly, if $w_j>0$, $j$ remains unchanged; otherwise, $j$ and $w_j$ are updated to the index and a copy of the weight of the next particle with a non-zero weight. This process is repeated until all pairs of samples have been iterated through.

The simulation's iteration interval $\Delta t$ must be adjusted based on the rate of stellar collisions. Among all collision pairs, we first calculate the mean value of the rates, $\langle R_{\rm col}\rangle$, and set the interval as
\be
\Delta t={\rm min}(f_{\rm rlx}t_{\rm rlx}, f_{\rm col}\langle R_{\rm col}\rangle^{-1}),
\label{eq:delta_t1}
\ee
where $f_{\rm rlx}=10^{-3}\sim 10^{-2}$ and $t_{\rm rlx}$ is the characteristic two-body relaxation time at the influence radius of the MBH ($M_\star(<r_h)=2\bh$). We set $f_{\rm col}=10^{-5}\sim 10^{-3}$, ensuring it is small enough for the convergence of the simulation results. The above time interval is adopted when the mass accretion is ignored. For the time interval when mass accretion of MBH is included, see Section~\ref{sec:evolving-case}.

\subsection{Testing the algorithm: A Plummer cluster}

Suppose there are two species, ``1'' and ``2'', with masses $m_1$ and $m_2$, number densities $n_1(r)$ and $n_2(r)$, and isotropic phase distributions $f_1(\vec v)$ and $f_2(\vec v)$, respectively. The total number of collisions, $N_{\rm col}$, between these species per unit radius and unit time is given by 

\be
\frac{d^2N_{\rm col}}{dr dt}=2\pi r^2 \int\int f_1(\vec v_1)f_2(\vec v_2)(|\vec v_1-\vec v_2|)R_{12}  d\vec v_1^{\,3} d\vec v_2^{\,3},
\ee

\noindent
where $R_{12}$ is similar to Equation~\ref{eq:gammaij} but with $i$ and $j$ replaced by $1$ and $2$. It has been demonstrated that, for a Plummer model~\citep{1911MNRAS..71..460P} with total mass $M$, characteristic radius $r_a$, and star number $N_\star=M/m_\star$, the collision rate per unit radius is given by~\citep{2002A&A...394..345F}:

\begin{align}
\frac{d^2N_{\rm col}}{dt dr}= 54\sqrt{2}\frac{\sqrt{G\rho_0}}{r_a}\frac{1}{\Theta_0^2} 
                          u^2(1+u^2)^{-21/4}\times \nonumber \\
[1+\Theta_0(1+u^2)^{1/2}],
\label{eq:collision_rate_plummer}
\end{align}

\noindent 
where 

\be\ba
u=\frac{r}{r_a},~~\rho_0=\frac{3}{4\pi}\frac{M}{r_a^3},~~\Theta_0=\frac{3}{\zeta N_\star}\frac{r_a}{r_\star},
\ea\ee

\noindent 
and $r_\star$ is the radius of a star.

The total rate of collisions is given by 

\be
\frac{dN_{\rm col}}{dt}=\frac{\sqrt{G\rho_0}}{\Theta_0^2}\left(\frac{17}{4}+\frac{26}{5}\Theta_0\right).
\label{eq:plummer_col}
\ee

Given $M=4\times10^6\msun$, $r_a=3.1$ pc, the theoretical expected value of the total collision rate is $\frac{dN_{\rm col}}{dt}=1.52\times10^{-7}$ yr$^{-1}$ ($\zeta=0.25$) and $\frac{dN_{\rm col}}{dt}=6.1\times10^{-7}$ yr$^{-1}$ ($\zeta=1.0$). 

We use \GNC~to calculate the collision rates for such a Plummer cluster, ignoring two-body relaxation. I.e. we only take into account stellar collisions and neglect the evolution of the nucleus' density, which would render it very difficult to compare the numerical results with the analytical equations. Figure~\ref{fig:plummer_test} shows the simulation results from \GNC. These results exhibit good consistency with the expectations from Equation~\ref{eq:collision_rate_plummer} {for different values of $M$, $r_a$ and $\zeta$}. {For example, for models with $M=4\times10^6\msun$ and $r_a=3.1$pc, the total collision rate obtained is $1.47\times10^{-7}$ yr$^{-1}$
(if $\zeta=0.25$) or $6.0\times10^{-7}$ yr$^{-1}$ (if $\zeta=1.0$), consistent with the theoretical expected rate from Equation~\ref{eq:plummer_col} within $2\%$.}
\subsection{Destructive and non-destructive events}
\label{sec:collision_product}
For simplicity, we limit our analysis to collisions between two main-sequence stars (MS) or between two red giants (RGs). To date, studies of stellar collision simulations have mainly focused on MS-MS collisions~\citep{1987ApJ...323..614B,2002ApJ...568..939L,2005MNRAS.358.1133F,2026ApJ..1000..162R}. Although other types of collisions are present, such as those involving white dwarfs~\citep{1989ApJ...342..986B}, RGs~\citep{1991ApJ...377..559R, 1999MNRAS.308..257B, 2009MNRAS.393.1016D, 2024MNRAS.tmp..388R}, no systematic studies have covered the vast parameter spaces of collision events involving various stellar types, stellar structures, impact parameters and relative velocities. Therefore, we ignore other types of collisions, such as those between stars and stellar-mass black holes. We prefer to proceed step by step, adding further layers of complexity only after addressing simpler idealizations.

We can determine if a collision event is destructive or non-destructive by estimating the escape velocity of the stellar progenitors of the outcome. The escape velocity for a collision between stars ``1'' and ``2'' is

\be
v_{\rm esc}=\sqrt{\frac{2(m_1+m_2)G}{r_{\rm hf,1}+r_{\rm hf,2}}},
\label{eq:vesc}
\ee

\noindent
where $r_{\rm hf,1}$ and $r_{\rm hf,2}$ represent the half-mass radii of the stellar objects.
According to~\citet{2005MNRAS.358.1133F}, approximately we have $r_{\rm hf,1}\simeq r_{\star,1}/4$ and $r_{\rm hf,2}\simeq r_{\star,2}/4$. In the case of equal-mass stars, this reduces to $v_{\rm esc}=2\sqrt{2m_\star/r_\star}$. {Figure~\ref{fig:mr} compares $r_{\rm hf}$ and $r_\star/4$ for a typical solar-like ($z=0.02$, $m=1 \msun$) star during its main-sequence (MS) and red-giant (RG) phases, as calculated with MESA~\citep[e.g.,][]{2019ApJS..243...10P, 2023ApJS..265...15J}. The figure demonstrates that our approximation $r_{\rm hf}\simeq r_\star/4$ holds well over both evolutionary phases of interest: $4r_{\rm hf}/r_\star=1.07-1.1$ during the MS phase, and $\sim 0.8-1.2$ during the RG phase with radii in the range $16\sim 60R_\odot$. }

As shown by the set of smoothed-particle hydrodynamics (SPH) simulations of~\citet{2005MNRAS.358.1133F}, destructive events frequently occur when $d/(r_{h,1}+r_{h,2})\le 1$ and $v_{\rm rel}\gtrsim v_{\rm des}\simeq 2v_{\rm esc}$ (See Figure 9 of ~\citet{2005MNRAS.358.1133F}). Hence, we define in this article destructive events as cases where 
\be\ba
v_{\rm rel}& \ge v_{\rm des}=2v_{\rm esc},~{\rm and}\\
d &\le 0.25(r_{\star,i}+r_{\star,j}).
\label{eq:vel_des}
\ea\ee 
Otherwise, we classify them as non-destructive. For $1\msun$ main-sequence stars and red giants with $\sim 1R_\odot$ and $25R_\odot$, respectively, the destructive infinity velocities are approximately $v_{\rm des}=4\sqrt{2m_\star/r_\star}\sim 2470\kms$ and $v_{\rm des}\sim 490\kms$, respectively.

In general, setting $0.25<\zeta\le 1$ in Equation~\ref{eq:gammaij} can cover more physical collisions between two stellar objects. We find that adopting $\zeta=1$ results in larger rates of non-destructive collisions, however, not affecting the rates of destructive collisions. As destructive events occur mostly frequently with small impact parameters (e.g., head-on collisions), we adopt $\zeta=0.25$ throughout this study, similar to those adopted in~\citet{2023ApJ...947....8A}. 
For destructive collisions, as we can see in Eq.~130 of~\citet{2023ApJ...947....8A}, the integration of the probability density is proportional to the surface of the star, contrary to collisions in globular clusters, where all impact parameters have the same probability. 

For pairs of collisions that are destructive, 
we update the weights of each particle by $\mathcal{W}\rightarrow {\rm max}(\mathcal{W}-N_{\rm col}(\Delta t),0)$. Particles with $\mathcal{W}$ values that become zero are removed from phase-space.

{According to the suit of about 15,000 SPH simulations of \cite{2005MNRAS.358.1133F} and~\citet{2026ApJ..1000..162R},
for each non-destructive event involving two MSs (MS-MS collisions), the merger can release gas ranging from a few percent to very significant fraction of the total masses. We assume that the resulting merger product has a mass of $0.9(m_1+m_2)$, with the remaining $0.1(m_1+m_2)$ released as free gas. We continue to track the dynamical evolution of these merging products.} 

{In contrast, destructive events result in the complete disruption of both stars. For MSs, we assume that all of their mass is released as free gas. For RGs, which have a compact core with $\sim 30\%$ of the total mass, we assume that the destructive event completely destroys the envelope and all their masses ($70\%$ of the total) are released as free gas. For simplicity, the remaining cores are no longer tracked and are removed from simulation.} The role and frequency of these destructive collisions are one of the key points we want to address in this study. 

The radius of the merged product is updated under the assumption that MS-MS mergers form new main-sequence stars, while RG-RG mergers form new red giants. In the simulation, such a merged particle is added to the pool of particles with the proviso that the event is non-destructive.

The above treatment of destructive collisions and merging events for both MS-MS and RG-RG collisions is obviously a crude approximation of what happens in a real galactic nucleus: the relative velocity required for destructive events may vary with the impact parameter and the masses of both stars; the properties of the merger product may depend on the detailed stellar structures and the configurations of the colliding orbits~\citep{2005MNRAS.358.1133F}. {We have also ignored the fly-by events, of which in reality may also release some fraction of gas during the encounter.} 

For MS-MS collisions, we will show later in Section~\ref{sec:scope_compare} that {our simplified treatment is already a good approximation of destructive events and merging events in general}. For RG-RG collisions, however, there is no previous reference that systematically covers their collisions. Nevertheless, we deem these approximations to be sensible to first order and interesting enough to address the question of how a self-consistent treatment of cluster evolution with a growing MBH affects the rates of stellar collisions. More sophisticated treatment of collision products from MS-MS and RG-RG collisions, as well as collisions between other stellar types, is beyond the scope of this work and is deferred to future studies. 

{For simplicity, we do not include stellar evolution as implemented in~\citet{2026ApJ...999..224Z}, since our simplified collision treatment is not currently applicable to collisions between arbitrary stellar types and a range of masses.}

\section{Self-Consistent Dynamics of Stellar Collisions}

\begin{table*}
    \caption{Models}
        \centering
    \begin{tabular}{|c|c|c|c|c|c|c|c|c|c|}\hline
      Name & $M_{\rm cl}^a$ ($\msun$) & $r_a^a$ & $\gamma^a$ &$r_{\rm eff,i}^b$ & $r_{\rm eff,f}^c$ & $r_{\rm h,f}^c$
       & $\bh^{d}$ ($\msun$) & Components$^{e}$ 
       & $M_{\bullet,f}$ ($\msun$)$^f$ \\  
    \hline
    M2    &  $4\times 10^7$  &$2.17$ & $1$ & $3.9$ & $6.1$  & $3.5$ & $4\times 10^6$, Fix
      & \multirow{3}{*}{\begin{tabular}{c} stars~($1\msun$) 
        \\SBHs~($10\,\msun$)\\ $f_\bullet=0.001$\end{tabular}}       & - \\ 
    \cline{1-8}\cline{10-10}
    M2G91    & $10^{9}$  & $2.56$ & $1$  & $4.6$ & $5.5$ & $1.9$  &\multirow{2}{*}{$10^4$, Growth} &    
          &  $5.3\times10^7$  \\
    \cline{1-7}\cline{10-10}
    M2G83    &  $4\times 10^7$  & $2.3$ & $1$  &  $4.1$    &  $5.5$ & $2.1$ &          &  &$1.8\times 10^6$ \\
    \hline
    M3    &  $4\times 10^7$  & $2.17$ & $1$ & $3.9$ & $6.4$ & $3.6$  & \multirow{3}{*}{$4\times 10^6$, Fix}
          & \multirow{7}{*}{\begin{tabular}{c} stars~($1\msun$) 
          \\SBHs~($10\,\msun$),$f_\bullet=0.001$ \\ RGs~($0.95\msun$),$f_{\rm RG}=0.02$\end{tabular}}       & $-$ \\ 
    M3\_2    &  $4\times 10^7$  & $2.17$ & $1.5$ & $2.6$ & $5.5$ & $3.4$ & 
            &  & $-$ \\ 
    M3\_3    &  $4\times 10^7$  & $2.17$ & $1.75$ & $2.1$ & $5.3$ & $3.3$ & 
            &  & $-$ \\        
    \cline{1-8}\cline{10-10}
        M3G91\_2    & $10^{9}$  & $2.56$ & $1.5$  & $3.3$ & $4.4$ & $2.0$ &\multirow{5}{*}{$10^4$, Growth} &    
              &  $8.6\times 10^7$ \\
		\cline{1-7}\cline{10-10}
		M3G91\_2(stc off)$^g$    & $10^{9}$  & $2.56$ & $1.5$  & $3.3$ & $5.3$ & $1.7$ &  &    
              &  $7.6\times 10^7$ \\
        \cline{1-7}\cline{10-10} 
        M3G83\_2  &  $4\times 10^7$  & $2.3$& $1.5$    & $2.9$    &  $5.0$  & $2.9$ &         & &$3.9\times 10^6$\\ 
        \cline{1-7}\cline{10-10}
        M3G83\_3  &  $4\times 10^7$  & $2.3$& $1.75$    & $2.3$    &  $4.8$  &  $3.5$ & &  &$5.1\times 10^6$\\ 
        \cline{1-7}\cline{10-10} 
        M3G71\_2    &  $2\times 10^{6}$  & $0.36$ & $1.5$  & $0.8$&  $5.9$    & $5.1$ &    &  &$2.1\times 10^5$\\ 
        \hline
    \end{tabular}
    %
        \tablecomments{         
In this table RG stands for red giants, SBH for stellar-mass black holes, and $M_{\bullet}$ the mass of the MBH.         
$^{a}$. $M_{\rm cl}$, $r_a$ (in units of pc), and $\gamma$ represent the initial mass of the cluster, the characteristic radius, and the slope index of the inner region of Dehnen's model, respectively.\\  
$^{b}$. $r_{\rm eff,i}$ (in units of pc) denotes the initial effective radius of the cluster.\\  
$^{c}$. $r_{\rm eff,f}$ and $r_{\rm h,f}$ (both in units of pc) represent the effective radius and the influence radius of the MBH, respectively, at $12$ Gyr.\\  
$^d$. The initial mass of the MBH. "Fix" indicates that the MBH mass is fixed to its initial value, while ``Growth'' indicates that the MBH can grow through loss-cone accretion (tidal disruption of stars, direct swallowing of stars falling into the event horizon, or SBHs with $r_p \leq 8r_g$) and accretion of gaseous material released by destructive stellar collisions.\\  
$^e$. $f_\bullet$ ($f_{\rm RG}$) represents the initial number fraction of SBHs (RGs) relative to the total number of stars.\\  
$^f$. The MBH mass at $12$ Gyr, as calculated by \GNC, for models that account for the mass growth of the MBH.\\
$^g$. Same as model M3G91\_2 but with the collision process turned off.\\
    \label{tab:model}}
    %
    \end{table*}

In this section we investigate the \textit{dynamics} of stellar collisions in NSCs and the \textit{evolution} of the corresponding event rates in two different fashions. In the first scenario, we assume that the mass of the MBH remains constant over all the story of the galactic nucleus, while in the second case we allow the MBH mass to grow due to both loss-cone accretion (as described in Paper II) and the accretion of gas released from both destructive and non-destructive collisions between stellar objects. Our objective is to find out how collision events appear in the self-consistent dynamics of NSC and a growing MBH.

The initial conditions for the models explored are presented in Table~\ref{tab:model}. We adopt an initial Dehnen profile~\citep{1993MNRAS.265..250D} for all models:
\be
\rho(r)=\frac{(3-\gamma)M_{\rm cl}G}{4\pi}\frac{r_a}{r^\gamma(r+r_a)^{4-\gamma}}.
\ee
\noindent
We consider two-component models (stars and stellar-mass black holes) and three-component models. For three-component models,  following~\citet{2023ApJ...947....8A}, we include an additional $2\%$ fraction of red giants with a mass of approximately $0.95\msun$.  
We assume all red giants have a radius of $25R_\odot\simeq 0.116$ AU, since this phase is where they spend most of their lives. For all models, we adopt an initial Dehnen density profile with either $\gamma=1.0$, $1.5$, or $1.75$.


\subsection{Non-evolving MBH masses}
\label{sec:non_evolving_cases}

\begin{figure*}
    \center
    \includegraphics[scale=0.75]{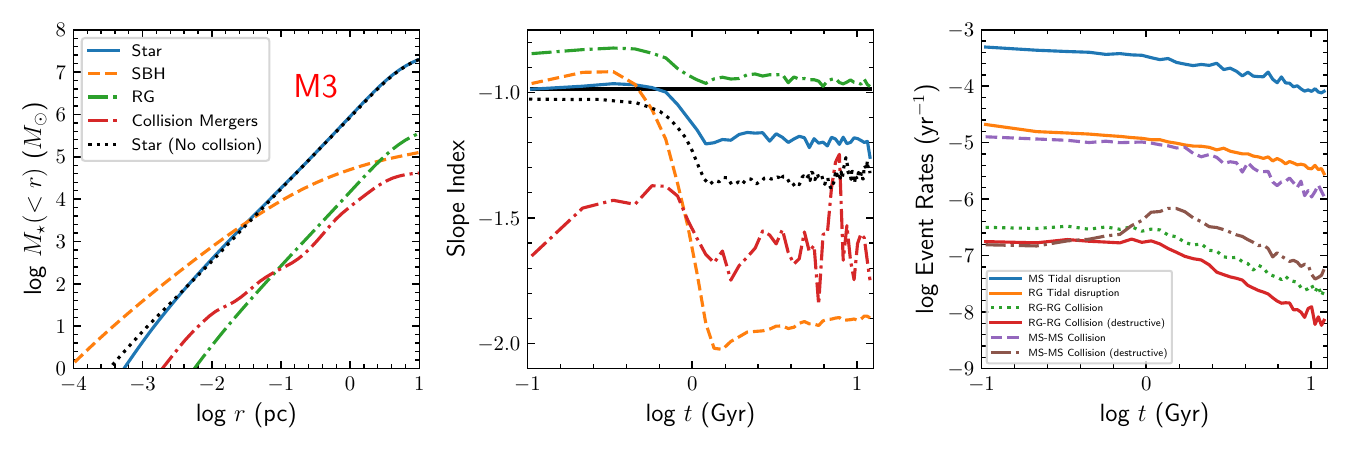}
    \includegraphics[scale=0.75]{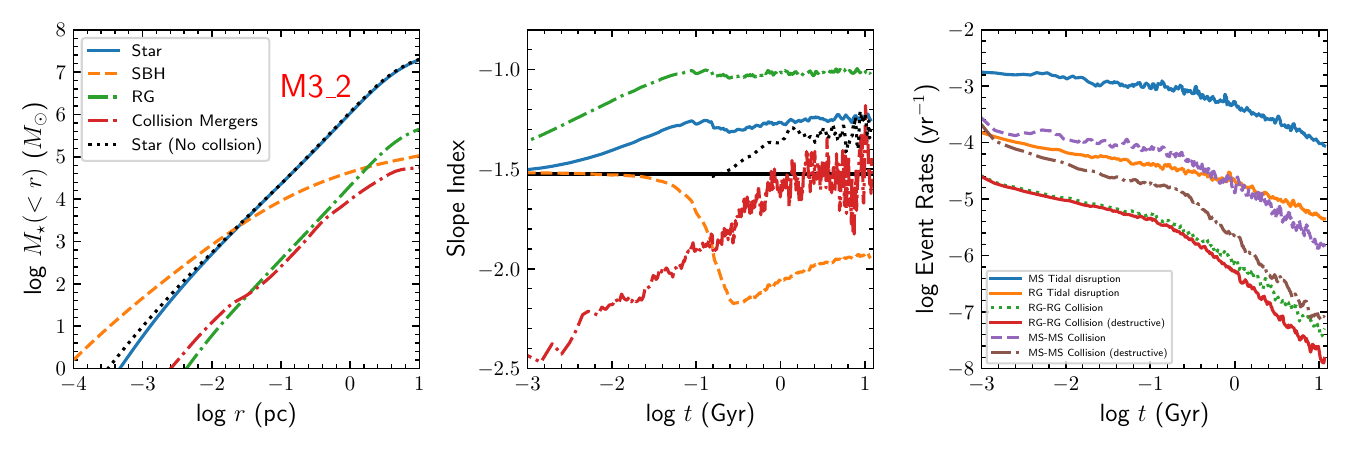}
    \includegraphics[scale=0.75]{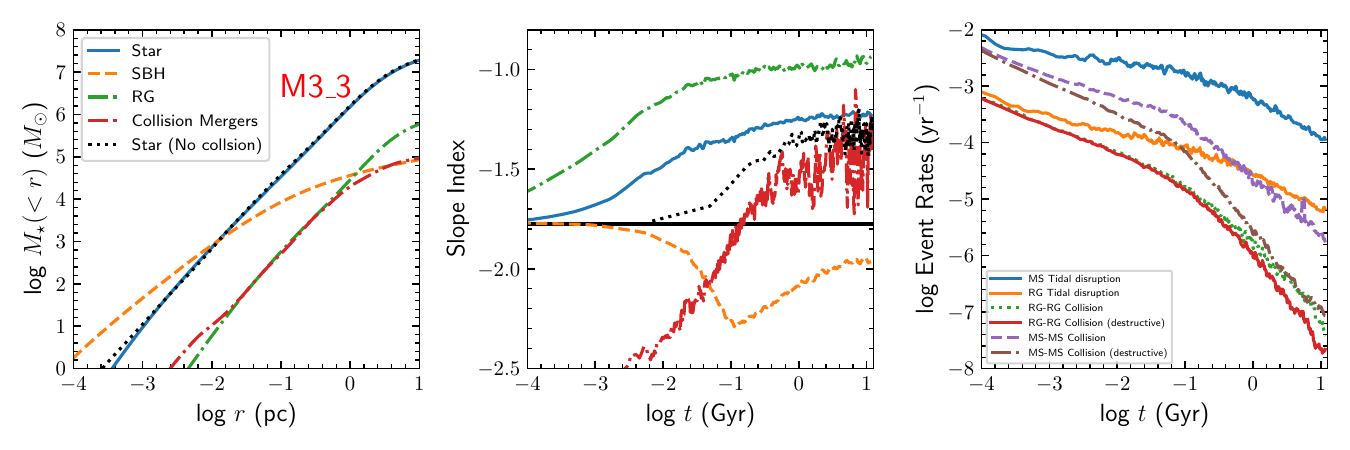}
    \caption{Left panels: Enclosed mass profile at $12$ Gyr for various models. The black dotted line represents the corresponding enclosed mass of stars if stellar collisions are ignored. Middle panels: Slope index of the density averaged at distances $10^{-3}$\,pc$<r<0.01$\,pc, with the line styles consistent with those in the left panels. The black solid line show the initial value of stars. Right panels: Evolution of the rates of tidal disruption events, total stellar collision rates, and rates of destructive collisions.}
\label{fig:col_evolution}
\end{figure*}

\begin{figure*}
    \center
    \includegraphics[scale=0.5]{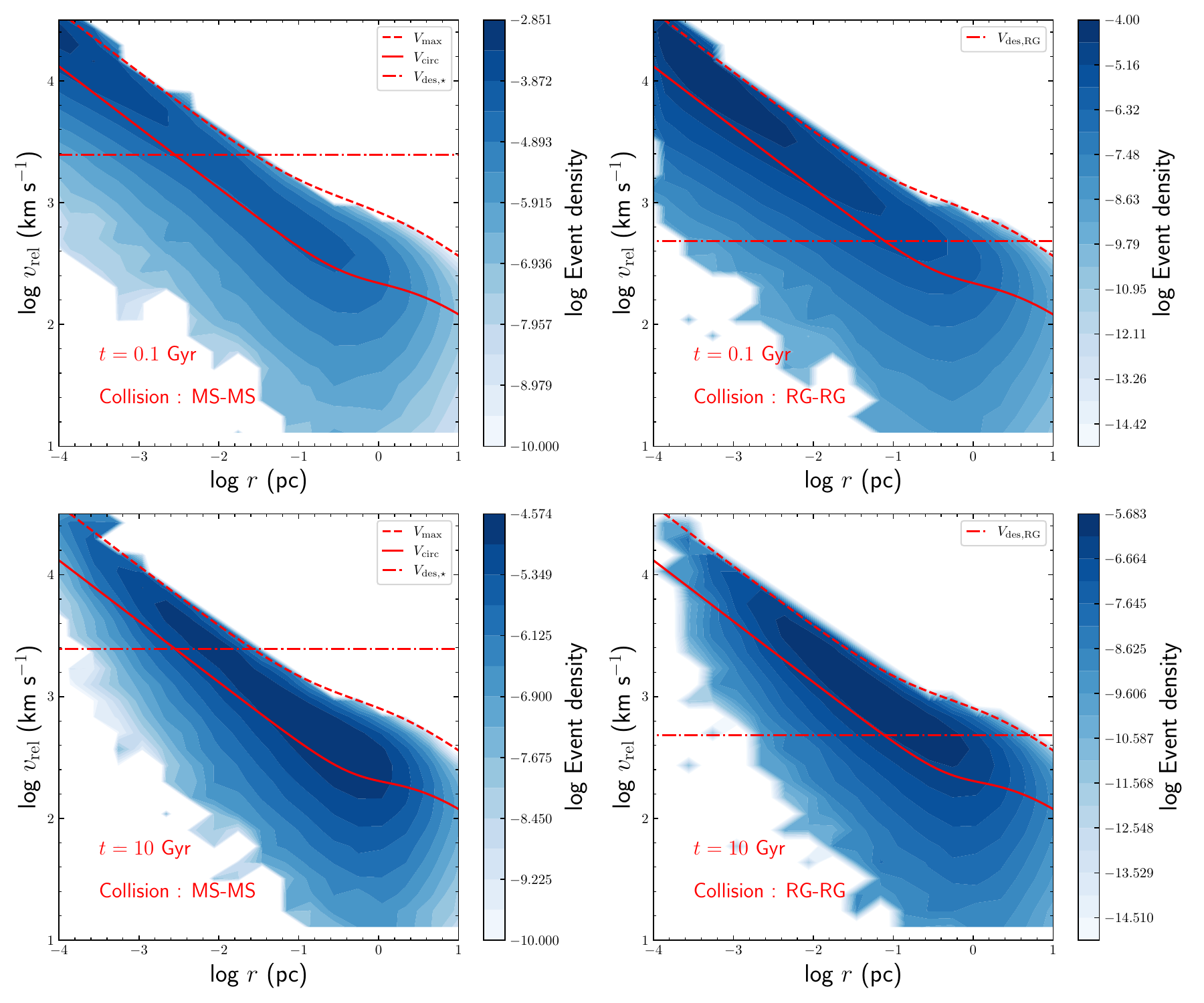}
    \caption{The spatial distribution of collision events in the $r-v_{\rm rel}$ space is shown at $t=0.1$ Gyr (upper panels) and $t=10$ Gyr (lower panels) for MS-MS collisions (left panels) and RG-RG collisions (right panels) in model M3\_2. {The colored contours show the number of collision events per unit logarithmic interval of $v_{\rm rel}/\sigma_0$ and   $r/r_0$, i.e., $d^2N/[ {\rm d}\log (v_{\rm rel}/\sigma_0) {\rm d} \log (r/r_0)]$, where $N$ is the number of collision events within the simulation interval, $r$ is the distance from the MBH, $r_0$ and $\sigma_0$ are the numerical units (See Section 2.1 of~\citetalias{2025ApJ...980..210Z}).} The red solid line represents the circular velocity $V_{\rm circ}=J_c/r_c$ {when considering both the potentials from the MBH and the stellar objects (See Equation 6 and 7 of~\citetalias{2025ApJ...980..210Z} for the detailed evaluation of $J_c$ and $r_c$)}. The red dashed line indicates the maximum relative velocity $V_{\rm max}=2\sqrt{\phi_\star(r)+G\bh/r}$, where $\phi_\star$ is the stellar potential. The red dash-dotted lines correspond to the destructive  velocity $v_{\rm des}$, above which defines destructive events, as given by Equation~\ref{eq:vel_des}.}
\label{fig:col_contour}
\end{figure*}

\begin{figure*}
    \center
    \includegraphics[scale=0.5]{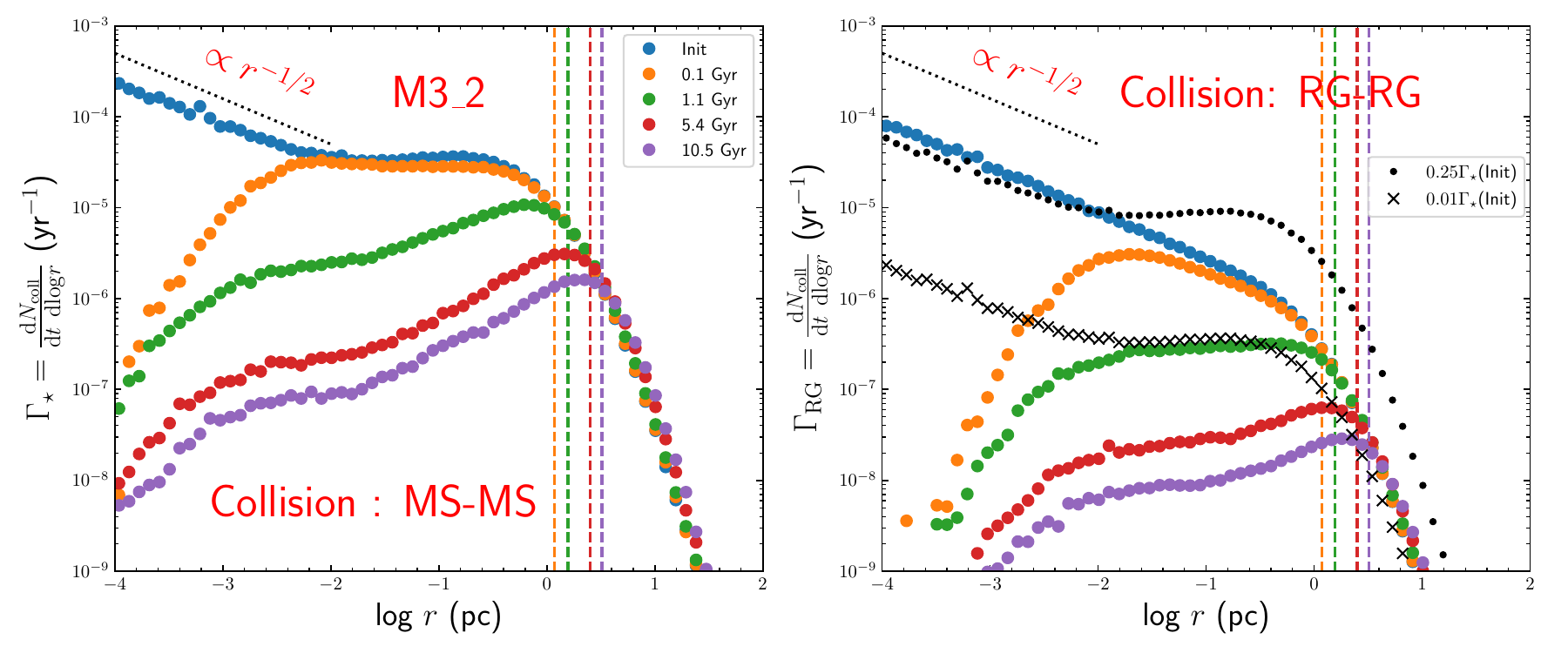}
    \includegraphics[scale=0.5]{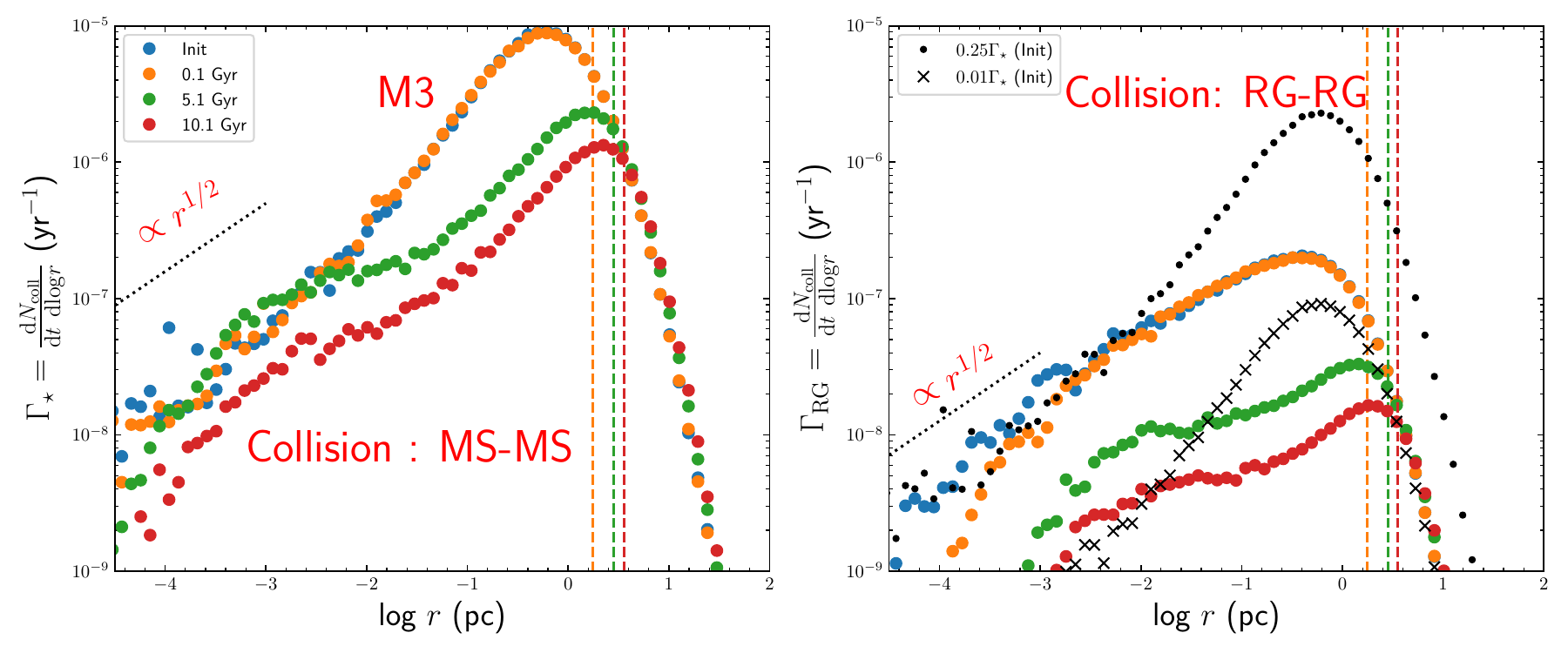}
    \caption{Distribution of collision events as a function of radial distance from the MBH is displayed. The left panels show results for MS-MS collisions, while the right panels correspond to RG-RG collisions. In the right panels, the black dots (crosses) represent the distribution of MS-MS collisions at the start of the simulation but are scaled down by factors of $4$ ($100$) for comparison with the RG-RG collision distributions. The dashed lines, in the same colors as the circles, indicate the position of the influence radius $r_h$, defined as $M_\star(r<r_h)=2\bh$.}
\label{fig:col_dstr}
\end{figure*}

We first investigate the dynamics of stellar collisions under the assumption that MBH mass is fixed; i.e. it does not grow over time. We perform simulations using \GNC\, for two-component models (model M2) and three-component models (model M3, M3\_2, and M3\_3, see their initial parameters in Table~\ref{tab:model}). The detailed results of dynamical evolution of stellar collisions are shown in the following sections.

\subsubsection{MS-MS Collisions}

Figure~\ref{fig:col_evolution} shows the dynamical evolution of density and stellar collision rates in models M3, M3\_2, and M3\_3. The evolution of density profiles and MS-MS collision rates in model M2 closely resembles those in model M3, as the RGs contribute only a small fraction to the total number and mass of stellar objects in the cluster. 

Comparing cases with and without stellar collisions, we see from the left panels of Figure~\ref{fig:col_evolution} that including stellar collisions alters the density profile of MS stars only within $r\lesssim 10^{-2}\sim10^{-3}$ pc. In regions $10^{-3}<r<10^{-2}$ pc, the slope of the density profile reduces from $-1.3\sim -1.4$ (without collisions) to $-1.2$ (with collisions). 

The middle panels of Figure~\ref{fig:col_evolution} demonstrate that both tidal disruption and stellar collision can deplete stars within $r<0.1$pc. If considering only the tidal disruptions (black dashed lines in all middle panels), the depletion occurs within $1$Gyr. However, when additionally the stellar collisions are included, the depletion timescale can shorten to be $\sim 0.01$ Gyr, especially for high initial density of the cluster in the inner regions (e.g. in models  M3\_3). At $12$Gyr, all three models converged to almost the same density profile. At regions $0.001$\,pc$<r<0.01$\,pc, density slopes are $-1.2$ for stars, $-2.0$ for SBHs and $-1.0$ for RGs.

The evolution of MS-MS collision rates in the first Gyr depends sensitively on the initial value of $\gamma$. In model M3\_3, which begins with a higher inner density ($\gamma=1.75$), the total MS-MS collision rate is up to $\sim4\times10^{-3}$ yr$^{-1}$ at the beginning of the simulation. However, the total rates rapidly decline, dropping by about three orders of magnitude within $\sim 1$ Gyr. For model M3\_2, which starts with $\gamma=1.5$, the initial rates are smaller, i.e., $\sim3\times10^{-4}$ yr$^{-1}$, but they also decline by about one order of magnitude. In these two models, the initial rapid decline of the collision rates is mainly attributed to the rapid depletion of stars in the inner regions through destructive stellar collisions.

For model M3 ($\gamma=1$), however, the collision rates remain almost constant in the first Gyr. The situation is slightly complicated by the effect of mass segregation. As shown in the top middle panel of Figure~\ref{fig:col_evolution}, the slope of the density profile in the inner regions ($r<0.01$ pc) steepens from $-1$ to about $-1.3$ within $\sim 0.5$ Gyr, due to both the mass segregation of SBHs and the relaxation of the particles' energy around MBH in the inner regions. The steepening of the density profile can even leads to an increase in the destructive collision rates in the first $\sim 0.5$ Gyr (See top-right panel of Figure~\ref{fig:col_evolution}). 

After the first Gyr, the collision rates of all models show a gradual decrease over time. This is mainly because the gradual expansion of the cluster and also the continuous depletion of stellar objects due to both the tidal disruption and the stellar collision. The expansion of the cluster is a result of the dynamical relaxation of the cluster (See more details in~\citetalias{2025ApJ...980..210Z}), which gradually decreases the mean density of the cluster, resulting in both smaller rates of tidal disruption events and collision events.

In both models M3\_2 and M3\_3, shortly after the simulation begins, the rate of destructive collisions becomes approximately an order of magnitude lower than the total rate. Thus, MS-MS collisions are quickly dominated by non-destructive events.  
{The above results suggest that very high collision rates of destructive events cannot be sustained for long periods due to the rapid depletion of stellar objects.}

MS-MS collisions have a high destructive velocity, $v_{\rm des}\simeq 2470\kms$. Figure~\ref{fig:col_contour} shows the distribution of collision events in the $r-v_{\rm rel}$ space for model M3\_2. At the beginning of the simulation ($t\sim 0.1$ Gyr), destructive collisions dominate, with most events having $v_{\rm rel}>v_{\rm des}$. However, as stars in the inner regions are consumed through collisions and tidal disruptions, the collision events become dominated by non-destructive events in the outer parts of the cluster, where most events have $v_{\rm rel}<v_{\rm des}$ (see the distribution at $t=10$ Gyr in the bottom left panel of Figure~\ref{fig:col_contour}).

The left panels of Figure~\ref{fig:col_dstr} show the evolution of the radial distribution of MS-MS collisions in models M3\_2 and M3. 
For destructive events of MS-MS collisions, the rate given by Equation~\ref{eq:gammaij} reduces to $R_{\rm col}\propto n(r) r^2_\star v_{\rm rel}$, where $r_\star$ is the radius of the star and $n(r)$ is the density as a function of radial distance $r$. In the inner regions, where $v_{\rm rel}\propto r^{-1/2}$ and $n(r)\propto r^{-\gamma}$, the rate of destructive events becomes:
\be\ba
\frac{d^2N_{\rm col}}{dt d\log r}\propto R_{\rm col} n(r) r^3 \propto r^{-2\gamma+5/2}r^2_\star.
\label{eq:col_theory}
\ea\ee
Thus, if $\gamma>5/4=1.25$, the total event rates are likely dominated by destructive events in the inner regions of the cluster. Otherwise, the total event rates are dominated by collision events in the outer parts of the cluster.

Initially, model M3\_2 has a slope index of $\gamma=1.5$. Thus, from Equation~\ref{eq:col_theory} it is expected that $d^2N_{\rm col}/dt d\log r\propto r^{-1/2}$, which agrees with the simulation results for $r\lesssim 0.01$ pc. Outside this radius, collision rates are dominated by non-destructive events.
Similarly, for model M3, which initially has a slope index of $\gamma=1$, Equation~\ref{eq:col_theory} predicts $d^2N_{\rm col}/dt d\log r\propto r^{1/2}$, which also matches the simulation results (see the bottom left panel of Figure~\ref{fig:col_dstr}).

After approximately 10 Gyr, the cluster reaches a quasi-relaxed state (though not a steady state, as the number of particles continues to decrease due to tidal disruption and stellar collisions). Due to the gradual depletion of stars, models M3 and M3\_2 both converge to a radial distribution of collisions dominated by events in the outer regions, peaking near $\sim 1$ pc. This peak location is close to the influence radius $r_h$, defined as $M_\star(r<r_h)=2\bh$. These results indicate that in a quasi-relaxed cluster, MS-MS collision rates are likely dominated by events occurring near the influence radius.

\subsubsection{Comparing with MS-MS collision product implemented from SCOPE}
\label{sec:scope_compare}
\begin{figure*}
    \center
    \includegraphics[scale=0.6]{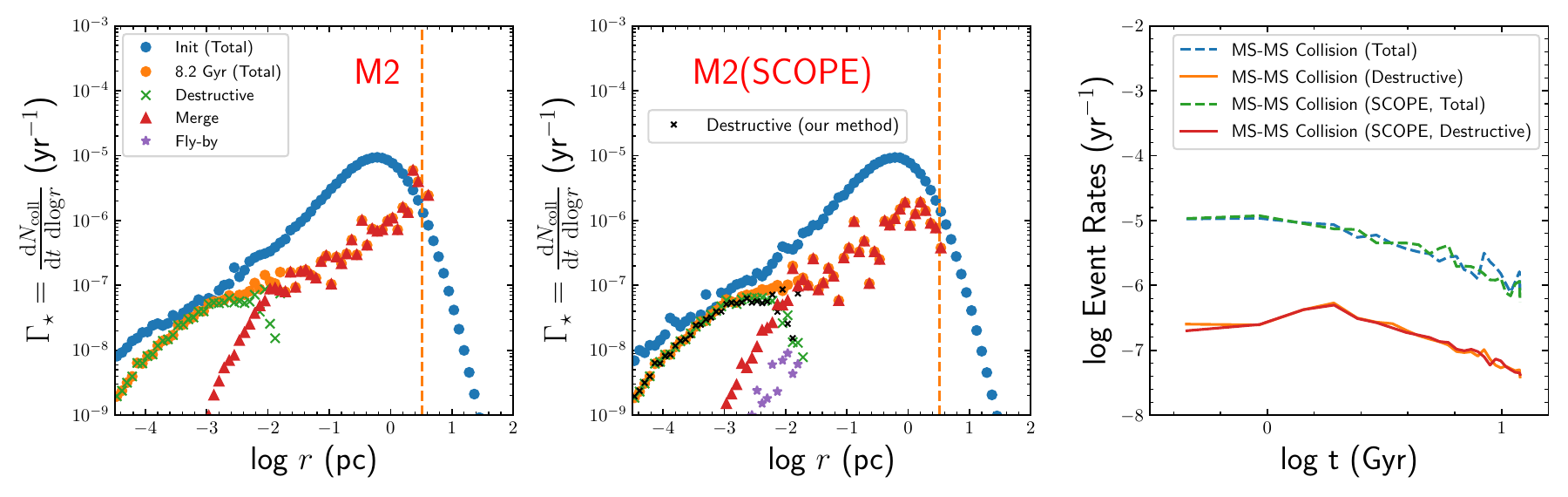}
    \caption{
    Comparing the collision results obtained using the criteria described in Section~\ref{sec:collision_product} with those adopting collision products from SCOPE~\citep{2026A&A...706A.315A}. The simulations adopt model M2 (see Table~\ref{tab:model} for its parameters). The left and middle panel show the distribution of collision rates for our method and for SCOPE, respectively. In both panels, the green crosses, red triangles and magenta stars  show the distributions of destructive events, merging events and fly-by events at $8.2$Gyr. In the middle panel, the black crosses show the same distribution of destructive events on the left panel, for the convenience of comparison. The right panel presents the evolution of the total and destructive collision event rates.
    }
\label{fig:col_m2_compare}
\end{figure*}
As described in Section~\ref{sec:collision_product}, whether a collision product is destructive or non-destructive depends simply on the relative velocity and the impact parameter. Since we consider only collision pairs of equal masses, according to~\citet{2005MNRAS.358.1133F}, this implementation can be considered roughly accurate. Nevertheless, we test the accuracy of our method using the outcomes from SCOPE~\citep{2026A&A...706A.315A}, a machine learning framework trained~\citep[scikit-learn,][]{Pedregosa2011} on a data sets of $\sim 16,000$ SPH simulations of main-sequence star collisions from~\citet{2005MNRAS.358.1133F}. In the test, for each colliding pair, we determine the product using SCOPE rather than the simple criteria of our implementation. The cluster adopts the two-mass component model M2 shown in Table~\ref{tab:model}. The results from SCOPE and our simple treatment of collision are compared in Figure~\ref{fig:col_m2_compare}. 

{We find that the results from our simple treatment are in general consistent with those from SCOPE, in terms of both the distribution and event rate of destructive collision events and mergers. The average mass loss for mergers in the SCOPE simulation is $\sim10\%$ of the total mass of the pair, which justifies our assumptions of merger mass loss described in Section 2.3.}

{Note, however, that our treatment of merging product remains approximate, as the fraction of released gas from mergers in the SCOPE simulation shows considerable scatterings, depending primarily on the impact parameter and relative velocity of the collisions.} In simulations adopting SCOPE, a small fraction of collisions can result in fly-by events, where both stars survive after the collision. Note that both simulations adopting our method and those from SCOPE consider only on near head-on collisions, i.e., adopting $\zeta=0.25$. Adopting $\zeta=1$ results in more merger events and fly-by events. However, the rates and distribution of destructive events are not affected.

\subsubsection{RG-RG collisions}

\begin{figure*}
    \center
    \includegraphics[scale=0.7]{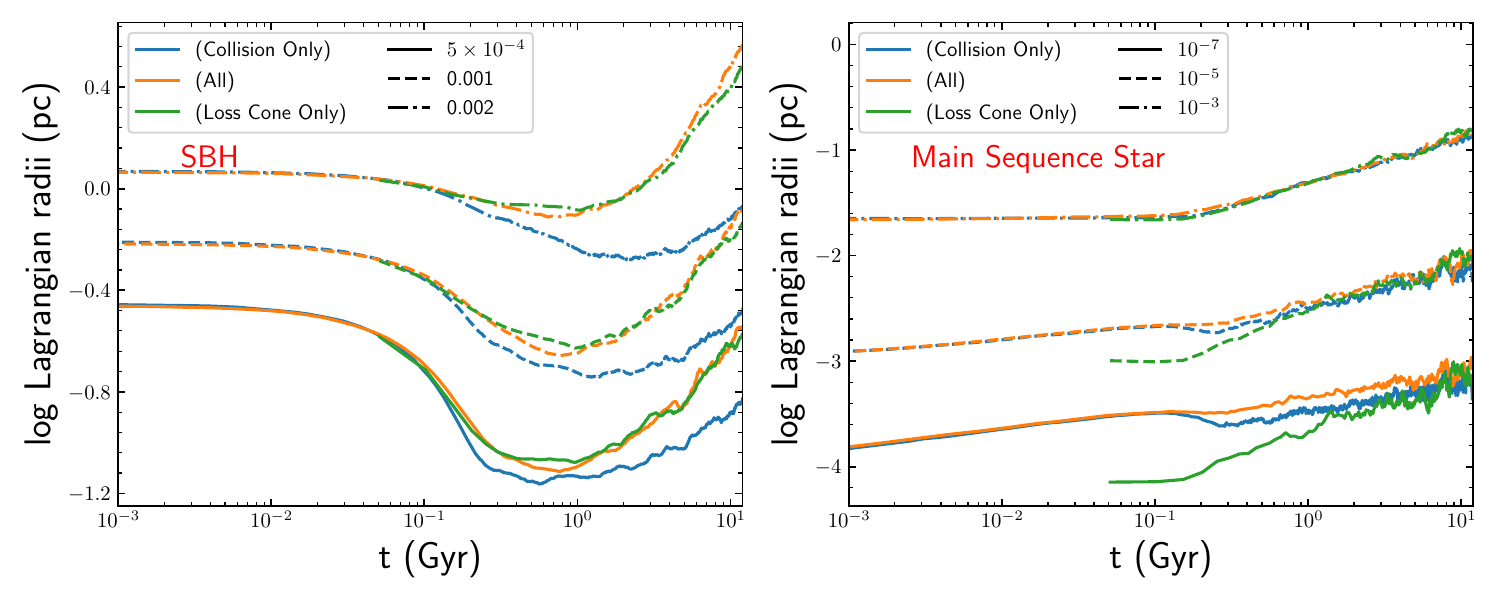}
    \caption{The evolution of Lagrange radii, within which $M(<r)/M_{\rm cl}$ corresponds to a specified value, is shown. Here, $M(<r)$ is the enclosed mass of SBHs or MS stars, and $M_{\rm cl}$ is the total mass of the cluster. The left panel presents the results for SBHs, while the right panel shows the results for MS stars. The fraction of enclosed mass corresponding to each line is indicated in the legends of each panel. The blue lines represent results when the effect of the loss cone is turned off, the green lines represent results when {stellar collisions} are excluded, and the orange lines represent results when both effects are included in the simulation.}
\label{fig:ge_m3_lag}
\end{figure*}

We find that, although normalized differently, the evolution of the density of RGs is very similar to that of MS stars. At 10 Gyr, the density profile of RGs is flatter than that of MS stars, as shown in the left and middle panels of Figure~\ref{fig:col_evolution}. 

In the inner regions of the cluster, the majority of collisions occur with high relative velocity $v_{\rm rel}\gg v_{\rm des}$ and the RG-RG collision is thus mostly destructive. 
According to Equation~\ref{eq:col_theory}, the destructive RG-RG collision rate per unit radius
is $\Gamma=d^2N_{\rm col}/(d\log r dt)\propto n^2 r_\star^2$, where $n$ is the 
density of particles. Given that RGs make up $2\%$ of MS stars 
and have a size 25 times larger, the destructive RG-RG collision event rates
is approximately $0.25$ of the MS-MS collision rate, i.e., $\Gamma_{\rm RG}=0.25\Gamma_{\star}$.

Conversely, in the outer regions of the cluster, where most collisions occur at low relative velocity $v_{\rm rel}\ll v_{\rm des}$, the RG-RG collision is most non-destructive 
($\Gamma\propto n^2 r_\star$) and the rate is expected to be $0.01$ of the MS-MS collision rate, i.e., $\Gamma_{\rm RG}=0.01\Gamma_{\star}$. Depending on whether the collisions are dominated by destructive or 
none-destructive events, the overall RG-RG collision rates across the cluster is about 
a fraction of $0.01\sim 0.25$ of those of MS-MS collision rates.

The blue dots in the right panels of Figure~\ref{fig:col_dstr} show the initial RG-RG collision rates in models M3 and M3\_2. It can be seen that the RG-RG collision rates, $\Gamma_{\rm RG}$, are approximately $0.25\Gamma_\star$ in the inner regions ($r\lesssim 0.02$ pc) of the cluster (see the black dots in the right panels of Figure~\ref{fig:col_dstr}) and about $0.01\Gamma_\star$ in the outer regions ($r\gtrsim 1$ pc) of the cluster (see the black crosses in the right panels of Figure~\ref{fig:col_dstr}). These findings are fully consistent with the analysis above. As a result, at the start of the simulation, the total RG-RG collision rate is approximately $0.04$ of the MS-MS collision rate in model M3 and $0.2$ in model M3\_2 and M3\_3.

From the right panel of Figure~\ref{fig:col_evolution}, the RG-RG collision rates are initially higher for clusters with higher inner densities, i.e., varies from $\sim 10^{-3}$ yr$^{-1}$ (for model M3\_3) to $\sim 10^{-7}$ yr$^{-1}$ (for model M3). After 10 Gyr of evolution, the rates of all models decrease to $\sim 10^{-8}$ yr$^{-1}$.

The right panel of Figure~\ref{fig:col_evolution} also demonstrates that a significant number of RG-RG collisions are destructive events. As shown in the right panels of Figure~\ref{fig:col_contour}, the escape velocity for RG-RG collisions is approximately $v_{\rm des}\sim 490\kms$, much smaller than that for MS-MS collisions. Even after 10 Gyr, a significant number of events still exhibit relative velocities above $v_{\rm des}$ (see the bottom-right panel of Figure~\ref{fig:col_contour}). This has important consequences, since destructive collisions will lead to interesting high-energy phenomena, as discussed in the work of \cite{2023ApJ...947....8A}.
\subsubsection{Collision Mergers}
Our simulations also result in a population of merge products from both the MS-MS and RG-RG collisions. As shown in Figure~\ref{fig:col_evolution}, after $12$Gyr, the enclosed mass of  these collision mergers are about a few times of $10^4\msun$ to $10^5\msun$. As these mergers have relatively larger masses ($\sim 2\msun$), their density profile are slightly steeper than those of RGs and MSs. As the models shown in Figure~\ref{fig:col_evolution} are Milky-Way-like NSCs, the results suggest that inside distance of S-stars ($r\lesssim0.04$pc~\citep{2009ApJ...692.1075G, 2017ApJ...837...30G}), {the enclosed total mass of mergers can be up to $\sim 100\msun$}. These merge products provide possible explanations of the G-gas clouds around the MBH~\citep{2014ApJ...796L...8W}.

However, note that the accuracy of the above results are limited by our simple implementation of the merging processes described in Section~\ref{sec:collision_product} and by our simplified NSC model that consists of three mass components. In reality, the NSC should contain multiple mass components, various stellar types and the merger products should depend on many details of the  stellar collisions. We defer the study of collision mergers under such a more comprehensive framework to future studies.

\subsubsection{Comparing the effects of loss cone and stellar collisions}

We compare the effects of loss cone dynamics and stellar collisions on cluster evolution by analyzing three possibilities: (1) only loss cone effects; (2) only stellar collisions; and (3) both effects combined. Note that in all three cases, particle depletion has considered when particles cross the lower- or upper energy boundaries~\citep[See Section 2.6 of ][]{2025ApJ...980..210Z}.

Taking model M3\_2 as an example, we depict the evolution of the Lagrange radii of stellar-mass black holes (SBHs) and MS stars in Figure~\ref{fig:ge_m3_lag}. From the solid lines in the right panel, we see that including collisions rapidly depletes stars in the inner regions of the cluster on a timescale of about 0.1 Gyr, as predicted in \cite{2023ApJ...947....8A}. During this timescale, the number of stars depleted by loss cone effects is negligible, and these effects do not significantly alter the stellar population in the very inner regions. However, over the interval $1\sim 12$ Gyr, the evolution of MS stars remains largely similar regardless of including loss cone effects or stellar collisions, except in the innermost regions of the cluster. For model M3 or M3\_3, we find that the depletion time of stellar density in the inner regions due to  stellar collisions varies from $0.01$ to $1$ Gyr.

For SBHs, the evolution of the density profile differs depending on whether these effects are included. While SBHs quickly diffuse in phase space towards the inner region within $\sim 1$ Gyr and then gradually expand outwards, loss cone effects lead to a significant depletion of SBHs in the inner regions. We find that MS-MS and RG-RG collision rates are nearly identical in configurations (2) and (3), indicating that the inclusion of loss cone effects does not significantly influence the rates of stellar collisions.

\subsection{A mass-growing MBH}
\label{sec:evolving-case}
\begin{figure*}
    \center
    \includegraphics[scale=0.95]{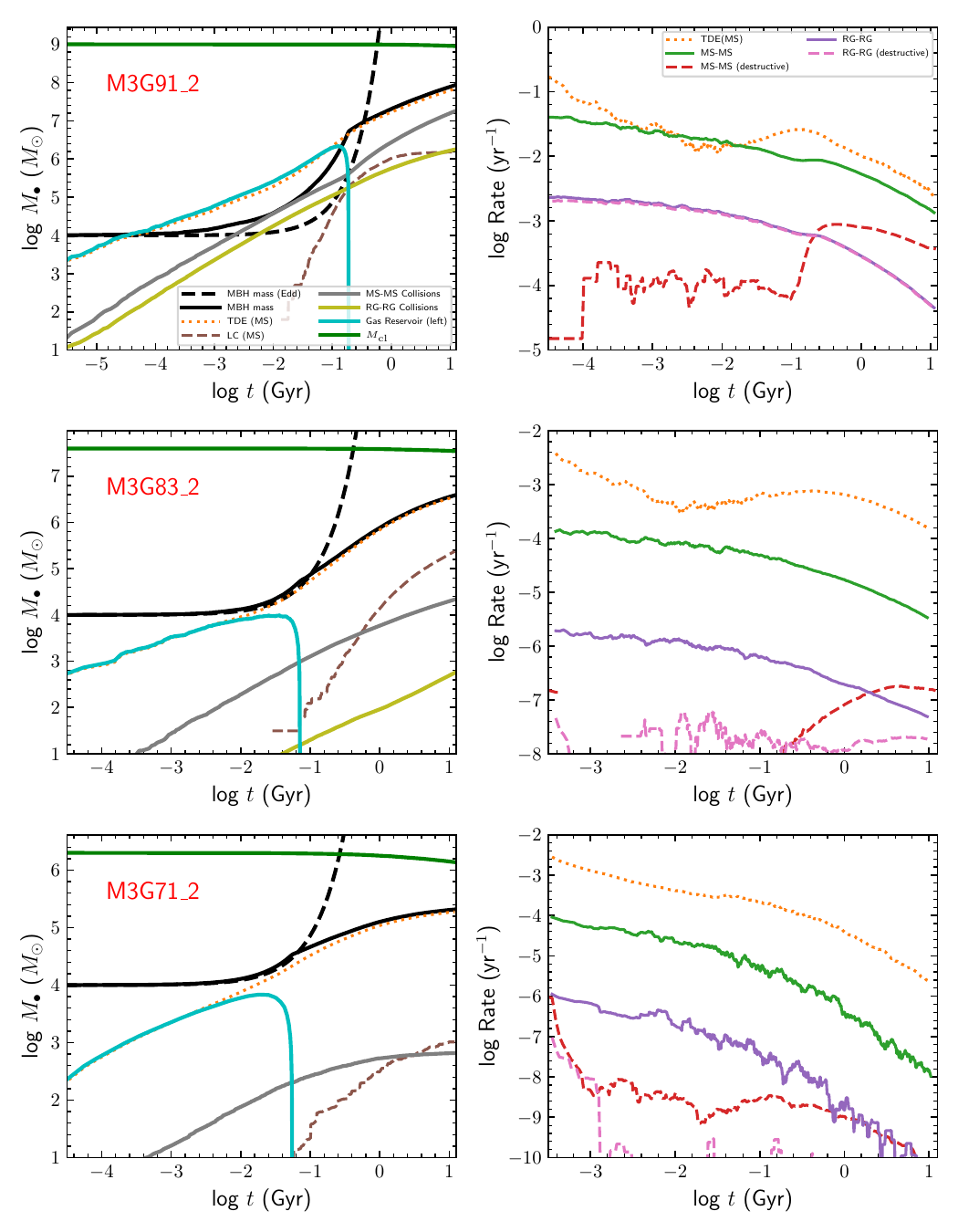}
    \caption{Left panels: The cosmological evolution of the MBH mass in a NSC due to loss cone accretion and the accretion of gas released by stellar collisions. Right panels: The cosmological evolution of tidal disruption event rates and stellar collision rates. In all panels, ``LC(MS)'' refers to the direct swallowing of MS stars if their pericenters lie within the event horizon of the MBH. ``TDE (MS)'' represents tidal disruption events of MSs. ``MBH mass (Edd)'' denotes the case where the growth of the MBH's mass is solely due to the accretion of gas from a reservoir of infinite mass under the Eddington limit. ``Gas Reservoir (left)'' indicates the remaining amount of gas in the reservoir after each iteration.}
\label{fig:col_evl_cos}
\end{figure*}

\begin{figure*}
    \center
    \includegraphics[scale=0.55]{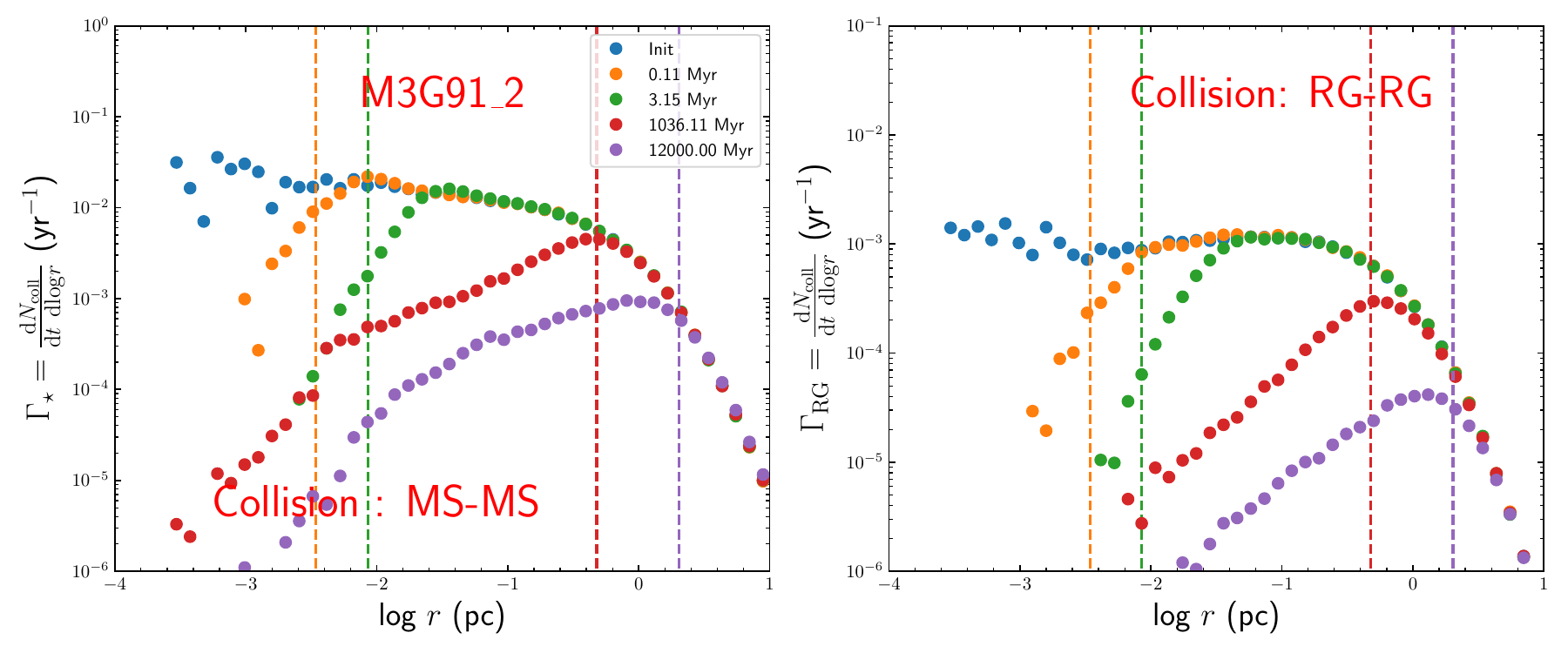}
    \caption{Similar to Figure~\ref{fig:col_dstr} but for Model M3G91\_2.}
\label{fig:col_dstr_M3G912}
\end{figure*}

\begin{figure*}
    \center
    \includegraphics[scale=0.55]{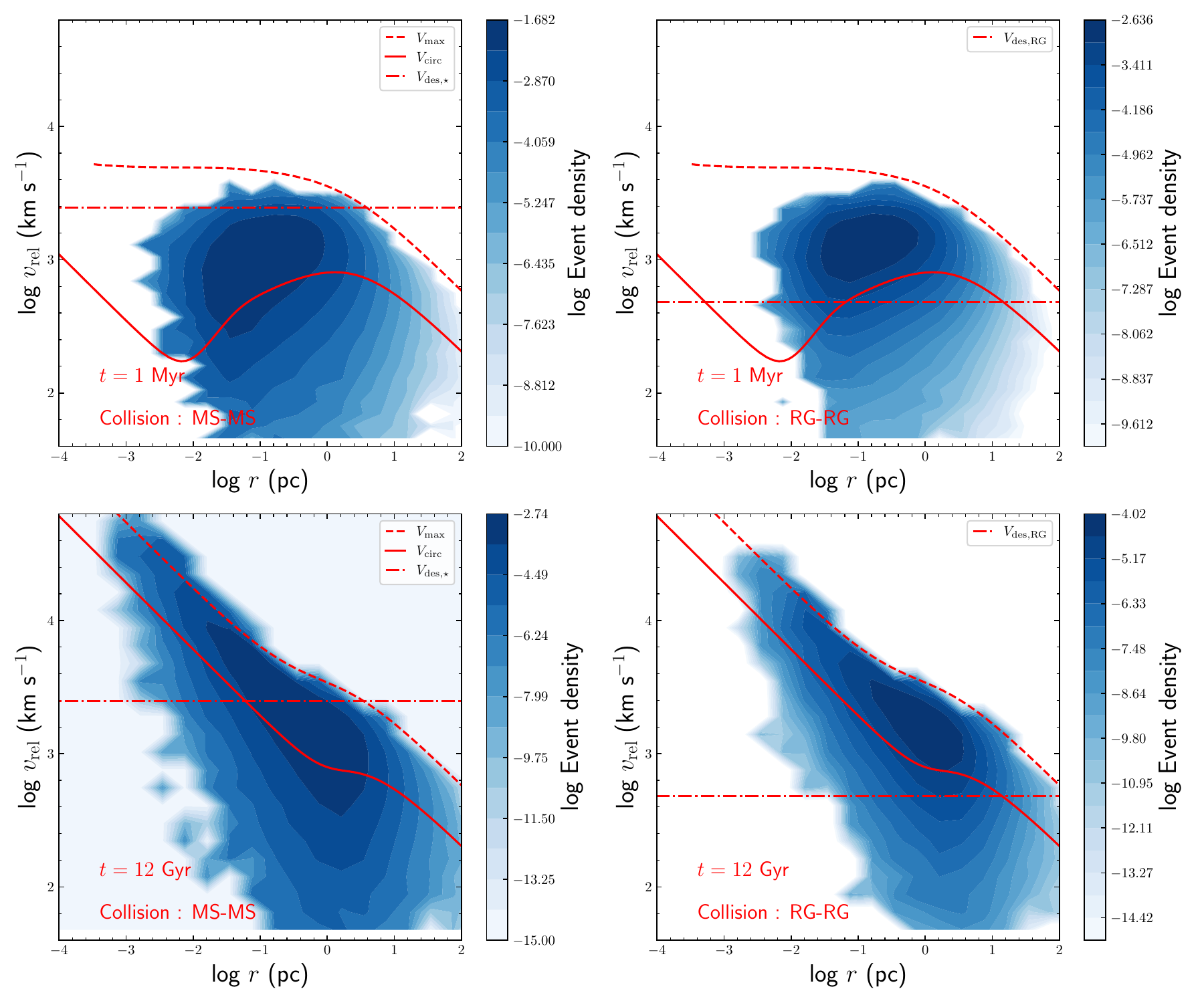}
    \caption{Similar to Figure~\ref{fig:col_contour} but for Model M3G91\_2.}
\label{fig:col_contour_M3G912}
\end{figure*}

After having evaluated the situation in which the MBH mass is ``frozen'' we now allow it to evolve, to grow, and we study the dynamics of stellar collisions. The MBH mass increases through accretion coming from both loss cone phenomena and gas released by collisional events. The growth of MBH mass via loss cone accretion, as described in~\citetalias{2025ApJ...980..210Z}, occurs by adding the entire mass of tidally disrupted stars to a gaseous reservoir, which is subsequently accreted by the MBH under the Eddington limit. Additionally, mass is added directly to the MBH if a star falls within the event horizon or if an SBH enters the innermost stable orbit ($r_p<8r_g=8\bh G/c^2$).

{We assume that a fraction $f_{\rm des}$ of the gas released during destructive events contributes to the gas reservoir, adding a mass of $\delta m=f_{\rm des}(m_1+m_2)$, where $f_{\rm des}=1$ for MS and $=0.7$ for RGs.  
For non-destructive events, the colliding objects merge to form a new MS/RG star, and a mass of $\delta m=0.1(m_1+m_2)$ is added to the reservoir. In the gas reservoir, we assume that $90\%$ of all these gas can be accreted by the MBH.}

As described in~\citetalias{2025ApJ...980..210Z}, when the gas accretion of the MBH is included, we limit the mass added by gas accretion in each simulation step to no more than $10\%$ of the current MBH mass. 
Thus, when $M_{\rm res}>0.1\bh$, in addition to $\Delta t$ given by Equation~\ref{eq:delta_t1},
the simulation time step is set to $\Delta t={\rm min}(\Delta t, t_{\rm acc})$, where $t_{\rm acc}=0.1\bh/\dot M_{\rm Edd}$ and $\dot M_{\rm Edd}=2.22\,\msun\,{\rm yr}^{-1}\bh/10^8\msun$ is the Eddington limit. 

We explore models similar to those in~\citetalias{2025ApJ...980..210Z}, in which the properties of NSCs after 12 Gyr of simulation are consistent or marginally consistent with the observed effective radii and cluster masses of NSCs in nearby galaxies~\citep{2016MNRAS.457.2122G}. To investigate RG-RG collisions, we also perform simulations for models that include RG components with masses of $0.95\msun$ in addition to the main-sequence and stellar-mass black hole components (the M3G model series in Table~\ref{tab:model}).

The cosmological evolution of the MBH mass and the event rates of various violent phenomena, including stellar collisions, are shown for several models from Table~\ref{tab:model} in Figure~\ref{fig:col_evl_cos}. Further details are provided in the subsequent sections.

\subsubsection{Events in massive NSCs}

For model M3G91\_2, which has a massive NSC with an initial mass of $M_\star=10^9\msun$, we find that the final MBH mass at $12$ Gyr is {$8.6\times10^7\msun$ when stellar collisions are included. Of this mass, $1.7\times10^7\msun$ is attributed to stellar collisions, while $6.8\times10^7\msun$ results from loss-cone accretion (i.e., tidal disruptions and the direct swallowing of stars).} If stellar collisions are excluded (model M3G91\_2 (stc off) in Table~\ref{tab:model}), the final MBH mass is reduced to $7.6\times10^7\msun$, which is accumulated primarily from tidal disruptions. These findings suggest that for massive clusters, including stellar collisions can increase the final MBH mass by approximately $13\%$. Although tidal disruption rates decrease, mainly due to the additional stellar depletion from collisions, a significant number of stars are converted into gas through destructive collisions, leading to greater overall MBH mass growth when collisions are included. 
 
For a similar model, M2G91, which initially has a flatter density profile ($\gamma=1$) than M3G91\_2, the final MBH mass is $\sim 5.3\times10^7\msun$. In this scenario, the MBH mass growth from stellar collisions is only $\sim 4.3\times10^6\msun$, approximately $24\%$ of that in model M3G91\_2.

The majority of the gas released during stellar collisions originates from MS-MS collision events, which are about an order of magnitude more frequent than RG-RG events. At the beginning of the simulation, the majority of MS-MS collisions are non-destructive. However, destructive fraction can be up to $30\%$ by $12$ Gyr, as the growing mass of the MBH triggers more destructive events. The present-day rates of both destructive MS-MS and RG-RG collisions can remain high. In model M3G91\_2, the rates are $3\times 10^{-4}$ yr$^{-1}$ and $4\times 10^{-5}$ yr$^{-1}$ for destructive MS-MS and RG-RG collisions, respectively.

Figures~\ref{fig:col_dstr_M3G912} and~\ref{fig:col_contour_M3G912} show the radial and radial-velocity distributions for collisions in model M3G91\_2. At the beginning of the simulation, collisions are concentrated in the inner regions of the cluster, where typical relative stellar velocities of collision events are a few times $10^3\kms$. This is because the cluster's gravitational potential is dominated by its stellar components, as the MBH has a relatively small initial mass ($\bh=10^4\msun$). Due to these low relative velocities, the majority of MS-MS collisions are non-destructive. In contrast, RG-RG collisions are predominantly destructive because of the lower escape velocities of the stars involved.

After $12$ Gyr of evolution, the MBH mass grows to near $10^8\msun$, which significantly increases the relative velocities for both MS-MS and RG-RG collisions, as shown in the bottom panels of Figure~\ref{fig:col_contour_M3G912}. Consequently, most collisions of both types become destructive. As shown in Figure~\ref{fig:col_dstr_M3G912}, these events occur predominantly within a radius that is a fraction of the MBH's influence radius. This figure further suggests that the location of these concentrated collision events evolves with the growing mass of the MBH.

\subsubsection{Events in Milky Way-like or smaller NSCs}

For NSCs comparable in size to the Milky Way's or smaller, the contribution of stellar collisions—both MS-MS and RG-RG—to the MBH mass growth is negligible. The vast majority of these collisions are non-destructive rather than destructive.

For a Milky Way-like NSC, such as in model M3G83\_2, the cosmological evolution of the MBH mass and the collision event rates are shown in the middle panels of Figure~\ref{fig:col_evl_cos}. Collisions contribute $\sim 2.1\times10^4\msun$ (MS-MS) and $560\msun$ (RG-RG), values that are much smaller than the contribution from tidal disruptions ($\sim 3.8\times10^6\msun$). If the inner density is increased, as in model M3G83\_3 with an initial $\gamma=1.75$, the contribution to MBH mass growth from collisions increases to $4.0\times 10^4\msun$ (MS-MS) and $\sim 1.2\times10^3\msun$ (RG-RG). In all of these cases, however, the mass growth of the MBH due to stellar collisions is relatively small.

In model M3G82\_2, the MS-MS collision rate starts at a high value of $\sim10^{-4}$ yr$^{-1}$, although nearly all of these events are non-destructive. This is because the relative velocities are low, as the MBH is small and the cluster's gravitational potential is shallow. By $12$ Gyr, the total MS-MS collision rate drops to $\sim 3\times 10^{-6}$ yr$^{-1}$, and few of these events become destructive as the MBH mass grows to $\sim 10^6\msun$. The RG-RG collisions begin with a rate of $10^{-6}$ yr$^{-1}$ that decreases to $\sim 4\times10^{-8}$ yr$^{-1}$ by $12$ Gyr.

For NSCs smaller than that of the Milky Way, such as in model M3G71\_2, the contribution from stellar collisions is only $\sim 600\msun$. The event rate for MS-MS collisions starts at $10^{-4}$ yr$^{-1}$ and gradually decreases to $10^{-8}$ yr$^{-1}$ by $12$ Gyr. The rates for RG-RG collisions are even lower. These results indicate that stellar collisions do not significantly contribute to MBH mass growth in low-mass NSCs and are rare events that are unlikely to be observed after $12$Gyr of evolution.

\section{Discussion and Conclusion}

In this study, we have investigated the dynamics of stellar collisions in nuclear star clusters (NSCs) and their interplay with the growth of the central massive black hole (MBH). We employed the \GNC~Monte Carlo code~\citep{ZA24a, 2025ApJ...980..210Z}, which we updated to self-consistently include the effects of stellar collisions alongside two-body relaxation and loss-cone physics. Crucially, our simulations track the growth of the MBH mass over cosmological timescales due to the accretion of stars (via tidal disruption events and direct swallowing) and the accretion of gaseous debris released during both destructive and non-destructive stellar collisions. We explored a range of NSC models, varying the cluster mass (from small to massive NSCs) and the initial inner density slope ($\gamma$).

Our main findings can be summarized as follows:

\begin{enumerate}
    \item {Impact on NSC Structure:} Stellar collisions alter the stellar density slopes in the innermost regions of the NSC ($r \lesssim 10^{-2}\sim 10^{-3}$ pc). We find that collisions  effects flatten the central density cusp, reducing the slope from initial values of $\gamma=1.5$ or $1.75$ to shallower profiles. This depletion of stars can occur rapidly if initially the density is high (on timescales of approximately $0.01$ Gyr), confirming analytical predictions regarding the scouring effect of collisions~\citep{2023ApJ...947....8A}. 

    \item {Evolution of Collision Rates:} The evolution of collision rates in the first Gyr is highly sensitive to the initial density profile. Systems starting with steep cusps exhibit very high initial collision rates (up to $\sim 10^{-2}\, \pyr$ for $\gamma=1.75$). However, the rapid depletion of stars causes these rates to decline sharply by several orders of magnitude within the first Gyr. After that, the self-consistent evolution of the cluster results in an expansion of cluster's sizes. Both the cluster's expansion and the depletion of stellar objects lead to a decreasing rate of stellar collision over cosmic time. In quasi-relaxed systems, the radial distribution of collision events tends to peak near the influence radius of the MBH.

    \item {Destructive vs. Non-destructive Collisions:} The nature of collisions evolves with the system. Red giant (RG) collisions are predominantly destructive due to their low destructive velocities ($v_{\rm des}\sim 490\kms$). Main-sequence (MS) star collisions ($v_{\rm des}\sim 2470\kms$) are often initially non-destructive, especially when the MBH mass is small and the cluster potential dominates. However, in models where the MBH grows significantly, the associated increase in stellar velocities shifts the balance, making MS-MS collisions more destructive over time.

    \item {Contribution to MBH Growth:} The contribution of stellar collisions to the MBH mass budget depends strongly on the mass of the host NSC.
    \begin{itemize}
        \item In massive NSCs (e.g., $M_{\rm cl}=10^9\msun$), stellar collisions provide an important fueling channel. The gas released by collisions contributed up to $\sim 1.7\times10^7\msun$ to the MBH growth, increasing the final MBH mass.
        \item In Milky Way-like or smaller NSCs, the contribution of stellar collisions to MBH growth is negligible (typically $< 10^4\msun$), with TDEs dominating the mass budget from stellar sources.
    \end{itemize}

    \item {Observational Implications:} The sustained high rates of destructive collisions in massive galaxies identify them as the most promising environments for observing the associated high-energy transients. In our most massive cluster models, the present-day rates of destructive MS-MS collisions remain high, up to $4\times10^{-4} \pyr$. In contrast, the rates in low-mass NSCs drop to negligible levels ($\lesssim 10^{-8} \pyr$) after $12$ Gyr of evolution.
\end{enumerate}

Our results demonstrate a link between stellar dynamics and MBH growth, particularly in the most massive galaxies. The feedback loop, where collisions fuel MBH growth, and the growing MBH enhances the velocity dispersion leading to more destructive collisions, plays an important role in the co-evolution of the system.

This study represents a significant step toward a self-consistent modeling of NSCs by incorporating the feedback from stellar collisions. However, our approach involves several simplifications. We assumed idealized stellar populations (fixed masses for MS and RGs) and relied on simplified prescriptions for collision outcomes based on historical SPH simulations~\citep{2005MNRAS.358.1133F}. While we verified that our MS-MS outcomes are consistent with those predicted by the machine-learning framework SCOPE~\citep{2026A&A...706A.315A}, future work should incorporate detailed stellar evolution and more sophisticated models for collision outcomes, including the effects of varying impact parameters and stellar structures. Furthermore, we neglected certain types of interactions, such as MS-RG collisions, which might lead to the stripping of the RG envelope, and collisions involving stellar-mass black holes. The inclusion of these processes may further affect the stellar populations and the availability of gas for accretion.

Despite these limitations, our findings underscore that stellar collisions are not merely a byproduct of dense stellar environments but active agents shaping the structure of NSCs, which can regulate the fueling of their central black holes in massive galaxies. The high rates of energetic, destructive collisions predicted for massive galaxies offer exciting prospects for transient astronomy, providing a unique window into the dynamics of the innermost regions of galactic nuclei.

\section{Acknowledgments}
\noindent
{We thank the anonymous referee for helpful comments and suggestions}. This work was supported in part by National Natural Science Foundation of 
China under grant No. 12273006. This work was also supported in part by the Key Project of the National Natural Science Foundation 
of China under grant No. 12133004. The simulations in this work are performed partly in the TianHe II National
Supercomputer Center in Guangzhou.


\begin{thebibliography}{1}


    
    
    
    
    
    \bibitem[Amaro-Seoane et al.(2017)]{AmaroSeoane17} Amaro-Seoane, P., Audley, H., Babak, S., et al.\ 2017, arXiv:1702.00786
    
    \bibitem[Amaro-Seoane(2018)]{2018LRR....21....4A} Amaro-Seoane, P.\ 2018, Living Reviews in Relativity, 21, 4. doi:10.1007/s41114-018-0013-8
    
    

	\bibitem[Amaro-Seoane et al.(2023)]{2023LRR....26....2A} Amaro-Seoane, P., Andrews, J., Arca Sedda, M., et al.\ 2023, Living Reviews in Relativity, Astrophysics with the Laser Interferometer Space Antenna, 26, 1, 2. doi:10.1007/s41114-022-00041-y

    \bibitem[Amaro Seoane(2026)]{2026A&A...706A.315A} Amaro Seoane, P.\ 2026, \aap, 706, A315. doi:10.1051/0004-6361/202557238

    \bibitem[Amaro Seoane(2023)]{2023ApJ...947....8A} Amaro Seoane, P.\ 2023, \apj, 947, 8. doi:10.3847/1538-4357/acb8b9

    
    
    \bibitem[Ashkenazy \& Balberg(2025)]{2025A&A...695A..98A} Ashkenazy, Y. \& Balberg, S.\ 2025, \aap, 695, A98. doi:10.1051/0004-6361/202453249

	\bibitem[Bailey \& Davies(1999)]{1999MNRAS.308..257B} Bailey, V.~C. \& Davies, M.~B.\ 1999, \mnras, 308, 257. doi:10.1046/j.1365-8711.1999.02740.x

    \bibitem[Balberg \& Yassur(2023)]{2023MNRAS.526.3688B} Balberg, S. \& Yassur, H.~B.\ 2023, \mnras, 526, 3688. doi:10.1093/mnras/stad2994

    \bibitem[Barausse et al.(2020)]{2020GReGr..52...81B} Barausse, E., Berti, E., Hertog, T., et al.\ 2020, General Relativity and Gravitation, Prospects for fundamental physics with LISA, 52, 8, 81. doi:10.1007/s10714-020-02691-1
    
    \bibitem[Benz \& Hills(1987)]{1987ApJ...323..614B} Benz, W. \& Hills, J.~G.\ 1987, \apj, 323, 614. doi:10.1086/165857

    \bibitem[Benz \& Hills(1992)]{1992ApJ...389..546B} Benz, W. \& Hills, J.~G.\ 1992, \apj, 389, 546. doi:10.1086/171230

    \bibitem[Benz et al.(1989)]{1989ApJ...342..986B} Benz, W., Hills, J.~G., \& Thielemann, F.-K.\ 1989, \apj, 342, 986. doi:10.1086/167656
    
	\bibitem[Dale et al.(2009)]{2009MNRAS.393.1016D} Dale, J.~E., Davies, M.~B., Church, R.~P., et al.\ 2009, \mnras, 393, 3, 1016. doi:10.1111/j.1365-2966.2008.14254.x

    \bibitem[Dehnen(1993)]{1993MNRAS.265..250D} Dehnen, W.\ 1993, \mnras, 265, 250. doi:10.1093/mnras/265.1.250
    
    
    \bibitem[Duncan \& Shapiro(1983)]{1983ApJ...268..565D} Duncan, M.~J. \& Shapiro, S.~L.\ 1983, \apj, 268, 565. doi:10.1086/160980
    
    
    
    
    
    
    \bibitem[Freitag(2003)]{2003ApJ...583L..21F} Freitag, M.\ 2003, \apjl, 583, L21. doi:10.1086/367813
    
    \bibitem[Freitag \& Benz(2001)]{2001A&A...375..711F} Freitag, M. \& Benz, W.\ 2001, \aap, 375, 711. doi:10.1051/0004-6361:20010706
    
    \bibitem[Freitag \& Benz(2002)]{2002A&A...394..345F} Freitag, M. \& Benz, W.\ 2002, \aap, 394, 345. doi:10.1051/0004-6361:20021142
    
    \bibitem[Freitag \& Benz(2005)]{2005MNRAS.358.1133F} Freitag, M. \& Benz, W.\ 2005, \mnras, 358, 1133. doi:10.1111/j.1365-2966.2005.08770.x
 
    
    \bibitem[Georgiev et al.(2016)]{2016MNRAS.457.2122G} Georgiev, I.~Y., B{\"o}ker, T., Leigh, N., et al.\ 2016, \mnras, 457, 2122. doi:10.1093/mnras/stw093
    
     
	\bibitem[Gillessen et al.(2009)]{2009ApJ...692.1075G} Gillessen, S., Eisenhauer, F., Trippe, S., et al.\ 2009, \apj, 692, 1075. doi:10.1088/0004-637X/692/2/1075

	\bibitem[Gillessen et al.(2017)]{2017ApJ...837...30G} Gillessen, S., Plewa, P.~M., Eisenhauer, F., et al.\ 2017, \apj, 837, 30. doi:10.3847/1538-4357/aa5c41

    
    \bibitem[H{\'e}non(1961)]{1961AnAp...24..369H} H{\'e}non, M.\ 1961, Annales d'Astrophysique, 24, 369
    
    \bibitem[H{\'e}non(1971)]{1971Ap&SS..14..151H} H{\'e}non, M.~H.\ 1971, \apss, 14, 151. doi:10.1007/BF00649201

    
    
    \bibitem[Hopman \& Alexander(2005)]{2005ApJ...629..362H} Hopman, C. \& Alexander, T.\ 2005, \apj, 629, 362. doi:10.1086/431475
    
    \bibitem[Hopman \& Alexander(2006a)]{2006ApJ...645L.133H} Hopman, C. \& Alexander, T.\ 2006, \apjl, 645, L133. doi:10.1086/506273
    
    \bibitem[Hopman \& Alexander(2006b)]{2006ApJ...645.1152H} Hopman, C. \& Alexander, T.\ 2006, \apj, 645, 1152. doi:10.1086/504400
    
	\bibitem[Jermyn et al.(2023)]{2023ApJS..265...15J} Jermyn, A.~S., Bauer, E.~B., Schwab, J., et al.\ 2023, \apjs, 265, 1, 15. doi:10.3847/1538-4365/acae8d 

    \bibitem[Lombardi et al.(2002)]{2002ApJ...568..939L} Lombardi, J.~C., Warren, J.~S., Rasio, F.~A., et al.\ 2002, \apj, 568, 2, 939. doi:10.1086/339060
    
    \bibitem[Marchant \& Shapiro(1980)]{1980ApJ...239..685M} Marchant, A.~B. \& Shapiro, S.~L.\ 1980, \apj, 239, 685. doi:10.1086/158155
    
    
    
    
    
    
        
    \bibitem[Murphy et al.(1991)]{1991ApJ...370...60M} Murphy, B.~W., Cohn, H.~N., \& Durisen, R.~H.\ 1991, \apj, 370, 60. doi:10.1086/169793
    
    \bibitem[Neumayer et al.(2020)]{2020A&ARv..28....4N} Neumayer, N., Seth, A., \& B{\"o}ker, T.\ 2020, \aapr, 28, 4
    
	\bibitem[Paxton et al.(2019)]{2019ApJS..243...10P} Paxton, B., Smolec, R., Schwab, J., et al.\ 2019, \apjs, 243, 1, 10. doi:10.3847/1538-4365/ab2241

    \bibitem[Pedregosa et al. (2011)]{Pedregosa2011} Pedregosa, F., Varoquaux, G., Gramfort, A., et al. 2011, Journal of Machine Learning Research, 12, 2825

    \bibitem[Plummer(1911)]{1911MNRAS..71..460P} Plummer, H.~C.\ 1911, \mnras, 71, 460. doi:10.1093/mnras/71.5.460
    
    \bibitem[Portegies Zwart \& McMillan(2002)]{2002ApJ...576..899P} Portegies Zwart, S.~F. \& McMillan, S.~L.~W.\ 2002, \apj, 576, 899. doi:10.1086/341798
    
    
    \bibitem[Quinlan \& Shapiro(1987)]{1987ApJ...321..199Q} Quinlan, G.~D. \& Shapiro, S.~L.\ 1987, \apj, 321, 199. doi:10.1086/165624
    
    \bibitem[Quinlan \& Shapiro(1990)]{1990ApJ...356..483Q} Quinlan, G.~D. \& Shapiro, S.~L.\ 1990, \apj, 356, 483. doi:10.1086/168856
    
	\bibitem[Rasio \& Shapiro(1991)]{1991ApJ...377..559R} Rasio, F.~A. \& Shapiro, S.~L.\ 1991, \apj, 377, 559. doi:10.1086/170385
    
    \bibitem[Rees(1988)]{1988Natur.333..523R} Rees, M.~J.\ 1988, \nat, 333, 523. doi:10.1038/333523a0

    \bibitem[Rizzuto et al.(2023)]{2023MNRAS.523.5439R} Rizzuto, F.~P., Naab, T., Spurzem, R., et al.\ 2023, \mnras, 523, 5439. doi:10.1093/mnras/stad1751

    \bibitem[Rose et al.(2026)]{2026ApJ..1000..162R} Rose, S.~C., Lombardi, J.~C., Gonz{\'a}lez Prieto, E., et al.\ 2026, \apj, 1000, 2, 162. doi:10.3847/1538-4357/ae459b
    
    \bibitem[Rose \& MacLeod(2024)]{2024ApJ...963L..17R} Rose, S.~C. \& MacLeod, M.\ 2024, \apjl, 963, 1, L17. doi:10.3847/2041-8213/ad251f

    \bibitem[Rose et al.(2023)]{2023ApJ...955...30R} Rose, S.~C., Naoz, S., Sari, R., et al.\ 2023, \apj, 955, 1, 30. doi:10.3847/1538-4357/acee75

    \bibitem[Rose et al.(2022)]{2022ApJS..260....2R} Rose, S.~C., Coughlin, E.~R., \& Zingale, M.\ 2022, \apjs, 260, 2. doi:10.3847/1538-4365/ac585d

    \bibitem[Ryu et al.(2024)]{2024MNRAS.tmp..388R} Ryu, T., Seoane, P.~A., Taylor, A.~M., et al.\ 2024, \mnras. doi:10.1093/mnras/stae396

    \bibitem[Sanders(1970)]{1970ApJ...162..791S} Sanders, R.~H.\ 1970, \apj, 162, 791. doi:10.1086/150709
 
    
    
    \bibitem[Shapiro \& Marchant(1978)]{SM78} Shapiro, S.~L., \& Marchant, A.~B.\ 1978, \apj, 225, 603
    
    
    \bibitem[Spitzer \& Hart(1971)]{1971ApJ...164..399S} Spitzer, L. \& Hart, M.~H.\ 1971, \apj, 164, 399. doi:10.1086/150855

    \bibitem[Spitzer \& Saslaw(1966)]{1966ApJ...143..400S} Spitzer, L. \& Saslaw, W.~C.\ 1966, \apj, 143, 400. doi:10.1086/148536

    \bibitem[Stone et al.(2017)]{2017MNRAS.467.4180S} Stone, N.~C., K{\"u}pper, A.~H.~W., \& Ostriker, J.~P.\ 2017, \mnras, 467, 4180. doi:10.1093/mnras/stx097
    
	\bibitem[Witzel et al.(2014)]{2014ApJ...796L...8W} Witzel, G., Ghez, A.~M., Morris, M.~R., et al.\ 2014, \apjl, 796, 1, L8. doi:10.1088/2041-8205/796/1/L8

    \bibitem[Zhang \& Amaro Seoane(2024)]{ZA24a} Zhang, F. \& Amaro Seoane, P.\ 2024, \apj, 961, 232. doi:10.3847/1538-4357/ad0f1a
    \bibitem[Zhang \& Amaro Seoane(2025)]{2025ApJ...980..210Z} Zhang, F. \& Amaro Seoane, P.\ 2025, \apj, 980, 2, 210. doi:10.3847/1538-4357/adaa7a
    \bibitem[Zhang \& Amaro Seoane(2026)]{2026ApJ...999..224Z} Zhang, F. \& Amaro Seoane, P.\ 2026, \apj, 999, 2, 224. doi:10.3847/1538-4357/ae3ca3


\end{thebibliography}
\end{document}